\documentclass{article}
\usepackage[utf8]{inputenc}
\usepackage{latexsym,float}
\usepackage{authblk}
\usepackage{epsfig}
\usepackage{epstopdf}
\usepackage{amsmath,hyperref}
\usepackage{amssymb}
\usepackage{graphicx}
\usepackage{eepic}
\usepackage{bbm}
\usepackage{textcomp}
\usepackage[T1]{fontenc}

\title{Dynamic properties of cyclic cosmologies}

\begin{document}
\author[1]{Petar Pavlović \thanks{petar.pavlovic@desy.de}}
\author[2]{Marko Sossich \thanks{marko.sossich@fer.hr}}
\affil[1]{\textit{\small{Department of Physics,
Ramashna Mission Vivekananda Educational and Research Institute,
Belur Math 711202, West Bengal, India}}}
\affil[2]{\textit{\small{University of Zagreb, Faculty of Electrical Engineering and Computing, Department of Physics,
	Unska 3, 10 000 Zagreb, Croatia}}}

\maketitle
\begin{abstract}
Our first goal in this work is to study general and model-independent properties of cyclic cosmologies. The large number of studies of bouncing cosmologies and different cyclic scenarios published recently calls for a proper understanding of the universal properties of cyclic models. We thus first review and further elaborate the common physical and geometrical properties of various classes of cyclic models and then discuss how cyclic Universe can be treated as a dynamic system. We then discuss how two theorems from dynamic systems analysis can be used to ensure the existence of cyclic cosmological solutions under certain conditions on the field equations. After this we proceed towards our second goal which is the application
of the obtained results to different frameworks of modified gravity theories: $f(R)$ gravity, dynamic dark energy and $f(T)$ gravity. We discuss the general requirements for the existence of cyclic solutions in these theories and also obtain various examples of cyclic cosmologies, while discussing their basic properties.  
\end{abstract}
\section{Introduction}
The idea that our Universe had an origin in the primordial singularity, usually denoted as the big-bang, is widely accepted both 
in the physical community, philosophy of science and popular science \cite{luminet, kragh, vilenkin, carrol, i}. However, this claim is lacking 
any empirical confirmation as well as any convincing theoretical justification. It is true that various independent observations, such as the 
abundances of chemical elements, growth of cosmological perturbations and microwave background measurements \cite{ellis1} are all consistent 
with the idea that the Universe evolved from an earlier state characterized by high temperatures and densities. However, such picture, also commonly called "the big-bang hypothesis" -- an ambiguity which further supports the confusion over the scientifically established opinion regarding the origin of the Universe, is in no way related to the question of the beginning of the Universe. This is due to the fact that the Universe can be evolving and changing its temperature, composition and properties even if it has no beginning or end. In the same sense, the fact that the Universe is expanding does not necessarily imply that the Universe needed to emerge from a single point, since it could as well be the case that the current state of expansion emerged from some earlier state of contraction, and not from a singular beginning.  On the other hand, a stronger reason for the physical existence of the big bang 
singularity is given by the singularity theorems of Hawking -- which show that, under the assumption of validity of general relativity and validity of the usual energy conditions for the matter-energy, there will always be geodesics which are geodesically incomplete, i.e. singularities necessarily need to appear on such spacetimes \cite{haw1,haw2,haw3}. But any direct application of these results to the early physics of our Universe is not justified, since it is precisely in this regime that we should assume that Einstein's general relativity will become invalid due to the quantum gravity effects. In fact, it is a well known result that even some very simple modifications of the field equations of general relativity -- which could effectively model the quantum corrections -- lead to non-singular solutions in which the big-bang is replaced with a bounce: a transition from an earlier phase of contraction to the expansion of the Universe \cite{bounce1, bounce2, bounce3, bounce4, bounce5, bounce6, bounce7, bounce8, bounce9, bounce10, bounce11, bounce12}. Therefore, everything that can actually be stated at this point is only that the early history of the Universe is still not known and that there are no actual reasons to assume that the Universe originated from a primordial singularity. Furthermore, such a sudden creation of something from nothingness would lead to familiar philosophical problems of creation $\textit{ex nihilo}$, and it would imply that the Universe essentially cannot be described by physics --  contrary to what has been proven by the development of science so far -- since at that point all the equations diverge. There are further reasons to suspect that Einstein's general relativity might perhaps not be a proper description of gravity even at energies much smaller than the ones characteristic for the Planck scale. The problem of the missing mass and missing energy density with negative pressure, which is stressed by many independent astrophysical and cosmological observations \cite{tamno1, tamno2, tamno3, tamno4, tamno5, tamno6}, is still neither solved nor properly understood after many decades of dedicated research. It is possible that these effects are not caused by some yet unobserved forms of matter and energy (called "dark matter" and "dark energy"), but are the consequence of incomplete validity of the equations of general relativity.\\ \\
For the stated reasons it is necessary to discuss physically viable models of the Universe which are free from the initial singularity, even if the proper theory of quantum gravity is still not known. In this respect it is natural to put a special emphasis on such theories which represent mathematical generalizations of general relativity, while keeping its fundamental physical principles preserved -- for such theories represent the most conservative first steps towards the new theory of gravity, and in the same time enable us to effectively introduce 
quantum corrections. We have already stated that different theories of such type can lead to a cosmological bounce. However, the bouncing picture does not describe the full evolution of 
the Universe, but just its transition from contraction to expansion -- and the question remains 
how did the Universe reach that state of contraction before the bounce. To say that the Universe simply started its existence and contraction from a special value of the scale factor leads to similar problems as the big-bang idea, and this does not answer the question which type of mechanism would actually cause its beginning in such a state. These issues are simply solved 
in the cyclic cosmology framework. In this paradigm, after the expanding phase which follows the bounce, the Universe undergoes a turnaround -- a transition from the expansion to contraction, subsequently leading to a new bounce and beginning of a new cycle. Since it is now known that the Universe is dynamic, the only consistent alternative to the idea of the Universe that had a beginning is the eternal Universe which undergoes an infinite number of phases of contraction and expansion. This type of cosmological scenario has many logical and physical advantages since it gives the natural and continuous evolution of the Universe without singularities, while in the same time solving additional problems such as the horizon problem (since the correlation between spacetime points can now be naturally established during the previous contraction cycle), and even the magnetogenesis problem, without any further assumptions and new theoretical ingredients \cite{nat}. \\ \\
During the years many different models of cyclic cosmology were developed \cite{cik1, cik2, cik3, cik4, cik5, 
cik6, cik7, cik8, cik9, cik10, cik11,cik13, cik14}. We have also recently proposed a rather general approach to cyclic cosmology, supported by the quantum inspired higher order curvature corrections to the standard Lagrangian of general relativity \cite{cik12}.  The problem is that all of the models need to assume some specific 
framework of modified/alternative theory of gravity. Moreover, many of them often use additional theoretical constructions to support the cyclical evolution (such as scalar fields and their couplings with gravitational sector, specific functional forms etc.) which mostly do not have any other theoretical justification or motivation, not to mention the empirical evidence. Thus the speculative assumptions taken in particular approaches considerably differ among each other, and the results obtained are therefore quite specific and to a high degree dependent on a chosen framework. Since we still don't know which, if any, of the alternative gravity models is preferred by Nature, it is very difficult to say which of the cyclic scenarios would properly describe the Universe in the case if it is indeed cyclic.\\ \\
The aim of this work will therefore be, for the first time according to our knowledge, to discuss general and model-independent properties of cyclic cosmologies. After proposing a simple mathematical framework suitable to describe different cyclic cosmological solutions, we obtain general results characterizing dynamic properties of cyclic cosmologies and then apply the obtained results to some concrete examples. We will particularly focus on more general frameworks, which can furthermore be motivated as effective approaches to quantum gravity -- such as dynamic dark energy, $f(R)$ and $f(T)$ gravity. We show that cyclic solutions naturally appear in all such theories of modified gravity if the certain mathematical conditions -- depending on the details of the field equations of the considered theory -- are satisfied.   \\ \\
This paper is organized as follows: in section II. we analyse general properties of cyclic cosmologies -- first by discussing the general geometrical properties of cyclic models in 2.1, and then approaching the cyclic universe as a dynamic system in subsection 2.2. In this subsection we introduce two general claims regarding the existence of cyclic solutions which are coming as a consequence of two important theorems regarding the existence of non-linear centers. In III. we study the application of the results obtained in II. to the case of modified $f(R)$ gravity, where we obtain some general properties of oscillatory solutions and also consider the specific example obtained by a reconstruction procedure. In IV. we discuss cyclic cosmological solutions that can be obtained in certain classes of dynamic dark energy models. Various necessary conditions for the realization of cyclic cosmologies in this context are discussed, as well as some concrete realizations. In V. we investigate the conditions for realization of cyclic cosmologies in $f(T)$ gravity. We show that cyclic solutions are possible due to the double-valued nature of the dependence between the scale factor and the Hubble parameter and analyse their dynamic properties in detail. In VI. we briefly discuss non-periodic oscillating cosmologies and we finally conclude in VII.

\section{General properties of cyclic cosmologies}
\subsection{Spacetime geometry of cyclic cosmologies}
In order to physically describe the eternal oscillating universe we assume the standard 
picture of the Universe as homogeneous and isotropic and given by the FLRW line element in spherical
coordinates:
\begin{equation}
 ds^{2}=-dt^{2}+a(t)^{2}\bigg[  \frac{dr^{2}}{1-kr^{2}}+ r^{2}(d\theta^{2} +  \sin^{2} \theta d\phi^{2})  \bigg],
 \label{metrika}
\end{equation}
where $a(t)$ is the scale factor and $k=\pm1$ describes the spatial curvature -- with $k=+1$ corresponding to positive spatial curvature, $k=-1$ negative curvature and $k=0$ leading to 
local flat space. In this work we concentrate on the flat Universe, $k=0$, since it appears to be favoured by observations during the current epoch \cite{planco}. Note that the observations in principle do not exclude the possibility that the Universe was not flat during the previous cycle. However, we will -- for simplicity and to avoid additional complications of the considered models -- keep the assumption of the Universe which is flat during its complete evolution. 
The content of the Universe is described as 
a perfect-fluid with the energy-momentum tensor:
\begin{equation}
 T_{\mu \nu}=(\rho + p)u_{\mu}u_{\nu} + pg_{\mu \nu},
 \label{idealni}
\end{equation}
where $\rho$ is the density, $p$ is the pressure, $u_{\mu}$ is the four-velocity which satisfies $u_{\mu}u^{\mu}=-1$. It is moreover assumed that the pressures and densities are related by the equation of state parameter, $w$, such that $p= w \rho$. 
The energy momentum conservation:
\begin{equation}
 \nabla_{\mu}T^{\mu \nu}=0,
\end{equation}
gives the equation for the change in the energy density:
\begin{equation}
 \dot{\rho}+3H(t)(\rho + p)=0,
 \label{conservation}
\end{equation}
where the dot is the time derivative and $H(t)=\frac{\dot{a}}{a}$ is the Hubble parameter. \\ \\
To study the evolution of cyclic models in general we propose to use the configuration space consisting of the following cosmological parameters: $a$ - giving the evolution of the physical distance between spatial points in dynamic Universe, $H$ describing the rate of expansion or contraction, and the Ricci curvature scalar, $R$, describing the curvature of the spacetime. The value of $R$ also describes the physical regime under study, since for high curvatures corresponding to the strong gravitational fields we expect that the proper theory of gravity departures from Einstein's general relativity due to quantum effects. We demand that in cyclic models these parameters, containing the full geometrical description of the Universe, always remain finite and well defined. For this to be possible the field equations of general relativity,$R_{\mu \nu} - (1/2)R g_{\mu \nu}=8 \pi G T_{\mu \nu} + \lambda g_{\mu \nu}$ (with $\lambda$ being the cosmological constant) need to be modified for strong gravitational fields as the result of quantum gravity corrections. From (\ref{metrika}) it then follows that the equations describing the Universe as an autonomous dynamic system in the space of parameters $a$, $R$, $H$ are:
\begin{equation}
\dot{a}=a H   
\label{skala}
\end{equation}
\begin{equation}
\dot{H}=\frac{1}{6}[R-12H^2] 
\label{habl}
\end{equation}
\begin{equation}
\dot{R}= g(a,H, R),    
\label{kurva}
\end{equation}
where $g(a,H,R)$ is some function given by the concrete theoretical framework in which the field equations of general relativity are modified. The only restrictive condition we take is that the alternative theory of gravity leads to the equation of the form (\ref{kurva}), which will indeed be satisfied for a large group of theories. 
In order that all quantities stay well defined in the cyclic universe, the scale factor needs to change from $a(t)=a_{min}>0$ to $a(t)=a_{max}$, while the trajectories representing the evolution of the system need to be periodic -- and thus given by closed orbits in the configuration space. An example of such cyclic universe is given in Fig 1.    
\begin{figure}[h]
\centering
    \includegraphics[width=0.9\linewidth]{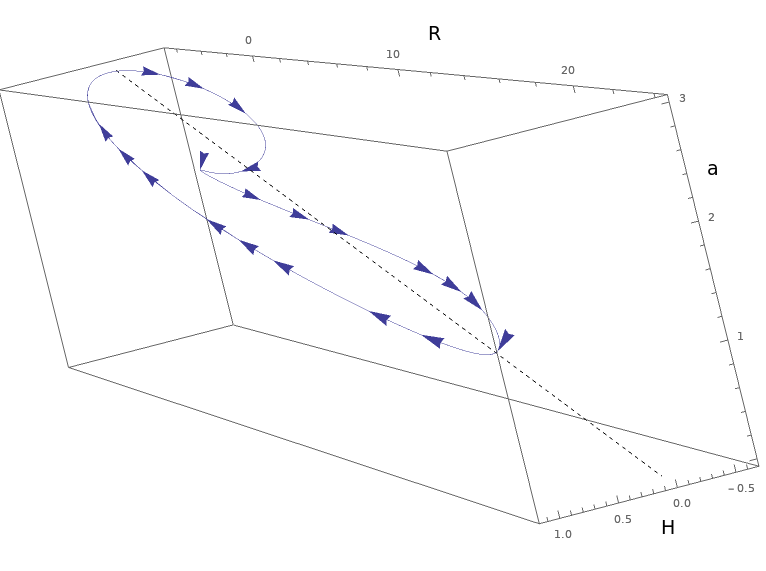}
       \caption{An example of cyclic universe in the configuration space given by the scale factor, the Hubble parameter and the Ricci curvature scalar. The arrows show the direction of the evolution of the Universe. The dotted line, corresponding to $H=0$, connects the bounce and the turnaround point of the cosmological evolution 
       } \label{Bfunr}
\end{figure}
\\ \\
The two essential points in the cyclic cosmologies are the bounce -- where the contraction of the later stage of 
the previous cycle is turned into the expansion in the new cycle, and the turnaround -- where the Universe enters from the expanding to the contracting phase. Both of the points are characterized by $H=0$. While in arbitrary cyclic cosmologies the number of such points, corresponding to transitions between expansion and contractions, could be arbitrary high, we restrict ourselves to the simplest case where there is exactly one bouncing and exactly one turnaround point. In this case, the line connecting those points and corresponding to $H=0$ (the dotted line in Fig.1) defines a plane which separates all the points on the trajectory in this configuration space to the ones corresponding to either expanding ($H>0$) or contracting ($H<0$) phase. It then also follows that 
the bouncing point corresponds to $a_{min}$, while the turnaround point corresponds to $a_{max}$. Since in Einstein's general relativity the bounce is replaced by a curvature singularity it is natural to infer that in cyclic models the value of the Ricci scalar would approach its maximum around the bounce. Since at the bounce $\dot{H}>0$ from (\ref{habl}) it follows that this value needs to be positive. If the maximum of $R$ is indeed reached during the bounce then it also follows that at the bounce point $\Ddot{H}=0$. The evolution of cyclic universe in general looks as follows. Every new cycle in the infinite history of the Universe 
begins from a high-curvature phase of cosmological bounce at which $R=R_{bounce}>0$, $H=0$ and 
$a=a_{min}$. The bounce is then followed by a phase in which $\dot{H}>0$, $\dot{a}>0$ and $\dot{R}<0$. This phase, like the bounce itself, needs to be based on the physics beyond general relativity and it corresponds to the violation of the effective null energy condition (while the null energy condition for the fluid components stays satisfied due to the higher-order corrections to standard general relativity). The viable models of cyclic cosmology also need to subsequently lead to such 
evolution which will be close to the one predicted by the $\Lambda$CDM model in the phases of 
radiation domination, matter domination and dark energy domination. After those phases the Universe needs to enter into the phase characterized by $\dot{H}<0$ and approach the turnaround point which is determined by $H=0$, $a=a_{max}$ and $R=R_{turnaround}<0$, where the last condition follows from equation (\ref{habl}). Note that 
the condition $\dot{H}<0$ also characterizes the previous $\Lambda$CDM phase, but unlike this previous phase, here we do not have any observational constraints on the cosmological evolution, so the possible evolution of the Universe approaching the turnaround point stays more general and subject only to the condition of decreasing Hubble parameter.  
\begin{figure}[h]
\centering
    \includegraphics[width=0.9\linewidth]{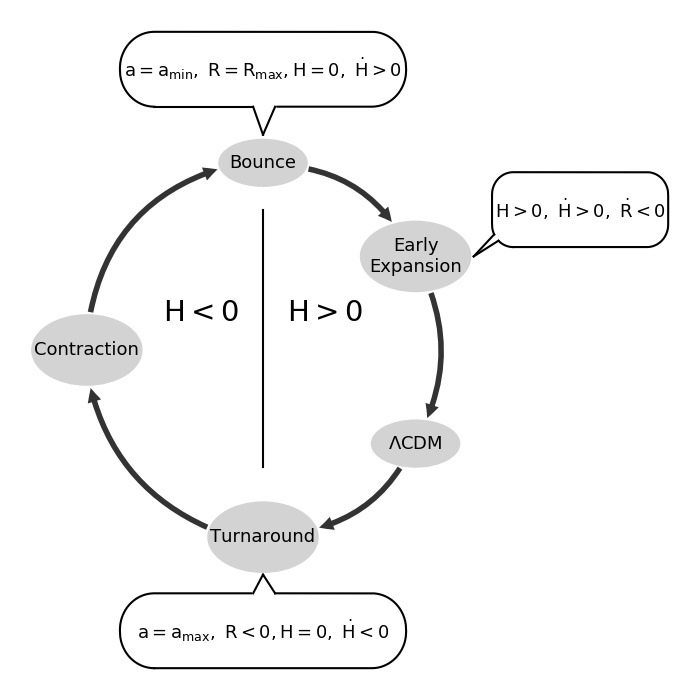}
       \caption{The evolution phases of viable cyclic cosmologies} \label{Bfunr}
\end{figure}
It is well known that the $\Lambda$ CDM paradigm, which introduces the small positive constant cosmological term, gives satisfactory description of the current accelerated expansion of the Universe. We can therefore assume that quantum gravity corrections of the lowest order can, at least at this scale, be effectively described as an effective cosmological term. However, such effective cosmological term introduced in the setting of modified gravity \cite{clifton} or based on the field theory considerations \cite{sola} will no longer be a constant, but will become a dynamic quantity. This will in general imply the change of the standard equation of state for cosmological constant: $p_{\Lambda }=- \rho_{\Lambda}$, so that the equation of state parameter will also in general become dynamic. If this approach is valid at least 
in the low curvature regime, then the corrected first Friedmann equation can be written as 
$H^{2}(t)=(8 \pi G/3) (\rho(t)_{rad}+ \rho(t)_{mat})+ \Lambda_{effective}(t)$. It then follows that the necessary condition for the turnaround is that the cosmological term changes from a small positive value, to a small negative value at the turnaround time, $t_r$, given by: 
\begin{equation}
    \lambda_{effective}(t=t_r)= - \frac{8 \pi G}{3}(\rho^{0}_{rad} a_{max}^{-4}+ \rho^{0}_{mat} a_{mat}^{-3}), 
\end{equation}
where $\rho^{0}_{rad}$ and $\rho^{0}_{mat}$ are the values of radiation and matter density today, and we have also assumed that the effective dark energy is not interacting with the energy-matter sector, so that the energy-momentum tensor for the matter and radiation stays conserved. The problem of phenomenological dynamics of the cosmological term, from the perspective of cyclic cosmology, was discussed in more detail in \cite{cik12}. Although the question about the dark energy dynamics is still unsolved from the point of view of current observations, the recent results suggest that the evolving dark energy does not contradict the measurements and even seems to be slightly preferred with respect to to the $\Lambda$CDM model \cite{dynamic1, 
dynamic2, dynamic3, dynamic4,dynamic5}. After the turnaround, the Universe enters into the contraction phase of its evolution, $H<0$, with the curvature scalar which eventually increases and approaches its maximal value, until the bounce is again reached and the new cycle begins. This general pattern of cyclic cosmological evolution is depicted in Figure 2. 
\subsection{Cyclic universe as a dynamic system}
The fact that the system (\ref{skala})-(\ref{kurva}) represents a set of autonomous differential equations enables us to use the methods of dynamic systems in order to understand the qualitative and global 
properties of its solutions. The application of dynamical systems methods in cosmology was rich and diverse in the past decades \cite{coll, novikov, copeland, ellis, coley, sebastian, biswas, 
mirza, saridakis, odin, alho, dunsby}, but according to our knowledge there was no work focusing on the analysis of cyclic cosmologies from this perspective. The central point of the dynamical system approach is to determine the fixed points of the considered system of differential equations, as well as their stability. The autonomous system of differential equations has the form
\begin{equation}
\dot{\mathbf{x}}=\mathbf{f(x)}   
\end{equation}
We can see that the system which we are considering, (\ref{skala})-(\ref{kurva}), has the appropriate form with
\begin{equation}
\mathbf{x}=\left(
\begin{array}{c}
 a \\
 H \\
 R \\ 
\end{array}
\right)
\end{equation} and 
\begin{equation}
\mathbf{f(x)}=\left(
\begin{array}{c}
 aH \\
 \frac{1}{6}(R-12H^{2}) \\
 g(a,H,R) \\ 
\end{array}
\right)
\end{equation}

The system (\ref{skala})-(\ref{kurva}) is very suitable for this approach in the light of cyclic cosmology, since from demanding that $a(t) \neq 0$, as discussed previously, it is very easy to see that the fixed point is given by $H^{*}=0$ and $R^{*}=0$, while the $a$ component of the fixed point is determined by the condition $g(a^{*},0,0)=0$, which needs to lead to the solution such that  $a_{min}<a^{*}<a_{max}$. The most common procedure for determining 
the type and stability of a fixed point is the linear stability theory -- in which the system is linearised around its fixed point. Therefore, in the expansion of a function defining a dynamical system, $\dot{\mathbf{x}}=\mathbf{f(x)}$, we consider only the first partial derivatives:
\begin{equation}
f_{i}(\mathbf{x}) \approx f_{i}(\mathbf{x^*}) + \sum_{j=1}^{n} \frac{\partial f_{i}(\mathbf{x^*})}{\partial x_{j}}(\mathbf{x} - \mathbf{x^*}), 
\label{exp}
\end{equation}
and stability of fixed points is thus encoded in the eigenvalues of the stability matrix, $J= \frac{\partial f_{i}(\mathbf{x^*})}{\partial x_{j}}$ evaluated at the fixed points. In our case the stability matrix is given by
\begin{equation}
J=  \begin{pmatrix}
    H       &    a    &   0   \\
    0       &    -12H    & \frac{1}{6} \\
    \frac{\partial g(a,H, R)}{\partial a}       & \frac{\partial g(a,H, R)}{\partial H}     &    \frac{\partial g(a,H, R)}{\partial R} 
   \end{pmatrix}
   \label{matrica}
\end{equation}
Those eigenvalues can be real or complex, and assuming that their real parts are all different from zero the linear stability theory will be sufficient to determine the stability of fixed points, according to Hartman–Grobman theorem. In this case, the fixed points can be classified on stable-nodes (if all of the eigenvalues have negative real parts), unstable nodes (if all of the eigenvalues have positive real parts) and saddle-points (if some -- but not all -- eigenvalues have positive real values, and others have negative real values). \cite{wiggins}. However, since we are discussing periodic cyclic cosmological solutions, all of these mentioned types of fixed points are not of interest to us, because their eigenvalues correspond to an attraction or repulsion from a considered fixed point along some direction in the phase space. On the other hand, we are interested in a very specific type of fixed points which are corresponding to closed orbits and are therefore neutrally stable -- in the sense that they neither attract nor repel nearby trajectories. Such fixed points are known as eliptic fixed points or centers and they correspond to purely imaginary eigenvalues \cite{wiggins}. Therefore, the requirement that the system (\ref{skala})-(\ref{kurva}) leads to cyclic cosmological solutions, corresponds to the condition
\begin{equation}
 \det[J - \lambda \mathbbm{1}]=\begin{bmatrix}
    -\lambda       &    a^*    &   0   \\
    0       &    - \lambda    & \frac{1}{6} \\
    \frac{\partial g(a,H, R)}{\partial a} |_{a^*,0,0}      & \frac{\partial g(a,H, R)}{\partial H}  |_{a^*,0,0}   &    \frac{\partial g(a,H, R)}{\partial R}|_{a^*,0,0} - \lambda
   \end{bmatrix} =0,  
   \label{linear}
\end{equation}
where $\lambda$ needs to be imaginary. The difficulty, however, here arises from the fact that the linear stability theory is inconclusive when it comes to the analysis of eliptic fixed points. Namely, such  fixed points have vanishing real parts and are therefore violating the conditions of the Hartman-Grobman theorem. This can be understood from the fact that eliptic fixed points or centers are not stable with respect to higher order non-linear corrections, which were neglected in (\ref{exp}), and which can perturb them into other types of fixed points. The presented condition is thus not sufficient and linear stability analysis on its own can lead to wrong conclusions. Further considerations are therefore necessary in order to discuss the conditions for the realisation of cyclic cosmologies. They are given by two important theorems related to our cosmological dynamical system (\ref{skala})-(\ref{kurva}). \\ \\
We state the following general claim (Claim 1): \textbf{Suppose that some gravity theory, defining a function $\mathbf{g(a, H, R)}$ in the equations for cosmological dynamics and leading to a continuously differentiable system  (\ref{skala})-(\ref{kurva})
has a following symmetry: $\mathbf{g(a, H, R)=-g(a, -H, R)}$. If $\mathbf{a^*}$, $\mathbf{H^*=0}$, and $\mathbf{R^*=0}$ is a center determined by the linear stability theory (i.e. given by the equation (\ref{linear})) then there exists a non-vanishing neighbourhood around this point such that all trajectories inside of it will correspond to symmetric cyclic cosmological solutions}. \\
To prove this claim we first note that, taking the stated assumption on the function $g(a,H,R)$, the system (\ref{skala})-(\ref{kurva}) is invariant under time inversion, $t \rightarrow -t$, if also $H \rightarrow -H$ and $R \rightarrow R$, while $a$ can by definition only stay positive. Such solutions are symmetric cyclic solutions with respect to the origin we choose to be defined by the bounce, $t_{bounce}=0$. Therefore, the system (\ref{skala})-(\ref{kurva})is reversible system in the sense that it has a reversing symmetry under time inversion. For reversible dynamic systems the existence of linear centers is sufficient to guarantee the stability of the 
center with respect to non-linear corrections, and therefore the existence of closed orbits around the fixed point \cite{center1, center2}. \\ \\
Such usage of the theorem for non-linear centers of reversible dynamic systems is obviously restricted to symmetric cyclic cosmologies. From the point of view of physically realistic cyclic models this requirement may be problematic, since symmetric models can lead to problems of instabilities and growing vector perturbations during the contracting phase \cite{bel,brande}, and non-symmetric models can be used to solve additional cosmological problems \cite{nat}. Therefore it is of interest to also have some other general criteria for the existence of cyclic solutions, not restricted to symmetric cosmological solutions. This can be achieved by using the theorem on the existence of closed orbits around the extreme point of the conserved quantity of the dynamical system. \\ \\
Claim 2: \textbf{Let us assume that the modified Friedmann equation in some theory of gravity, with the matter content of the Universe given by $n$ different components of energy density, $\mathbf{\rho_{i}}$, associated with the equation of state parameter for each component given by $w_i$, 
takes the form $\mathbf{F(a,H, R)=}\sum_{i}^{n} \mathbf{\rho_{i}(t)}$. 
Then the conserved quantity is given by 
\begin{equation}
    I(a,H,R)=\sum_{i=1}^{n}F(a,H,R)a^{3(1+w_i)}- \sum_{\zeta=1}^{n}\Big( \sum_{i=1,\zeta\ne i}^{n}\rho_{i}^{0}a^{-3(w_i-w_{\zeta})}\Big).
\end{equation}{}
If $I(a,H,R)$ has a strict local extremum at the fixed point given by $\mathbf{a^*}$, $\mathbf{H^*=0}$, and $\mathbf{R^*=0}$ satisfying (\ref{linear}), then there exists a non-vanishing neighbourhood around this point such that all trajectories inside of it will correspond to symmetric cyclic cosmological solutions.} Note that in the case of the Universe filled only with dust and radiation, the conserved quantity is simply given by $I(a,H,R)=F(a,H,R)(a^4+a^3)- \rho_m^{0} a -\rho_{r}^{0}/a$. To prove this claim we note that by (\ref{conservation}), by using $p=w \rho$, the quantity $I(a, H, R)$ constructed as the sum of all contributions corresponding to energy densities at a given moment, $\rho_{i}^{0}$, will stay conserved on cosmological trajectories in the configuration space. 
 Therefore, $I(a, H, R)$ is a first integral of autonomous system (\ref{skala})-(\ref{kurva}) in the sense that it is constant on solutions of this system. Now we can use the dynamical systems theorem which guarantees that if a point $(a^{*},H^*, R^*)$ is a strict local extremum of a first integral of the autonomous system of differential equations 
 then this point is a stable equilibrium point of the system, and thus the center of this system will be stable \cite{center2}. \\ The importance of this claim also comes from the fact that, even in the case where linear theory does not predict a center, it can be determined that the fixed point is stable if the considered integral $I(a,H, R)$ has a strict local extremum there. This is of general interest for dynamical analysis of cosmological equations, as the discussion on nature of fixed points in three or more dimensions with vanishing real parts of eigenvalues can otherwise become quite complex.\\ \\
 As there are many possible trajectories in the phase space, corresponding to different 
 gravitational theories and initial conditions, it is of interest to 
 somehow compare the physically relevant quantities characterizing specific models. One set of such parameters is given by the characteristic values of coordinates during the bounce ($a_{min}$, $H_{bounce}=0$, $R_{bounce}$) and turnaround ($a_{max}$, $H_{t_r}=0$, $R_{t_r}$). This type of information is of course only local and describes only the two most important points of cosmological evolution. 
 A global type of characterization of various cyclic cosmologies is 
 given by the integral in the configuration space, which is proportional to the period of the cyclic Universe
 \begin{equation}
\Gamma= \int_{a_{min}}^{a_{max}} \int_{H_{min}}^{H_{max}} \int_{R_{min}}^{R_{max}} da dH dR     .
 \end{equation}

\section{Cyclic cosmologies in $f(R)$ modified theory of gravity}
\subsection{A short review of $f(R)$ gravity}
One of the first attempts to modify Einstein's General
Relativity was simply to change the Einstein-Hilbert
action to a new more general action as a function of
curvature preserving all the symmetries of a viable
General Relativity. The action is given by  \cite{Buchdahl1}:
\begin{equation}
\mathcal{S}= \frac{c^{4}}{16 \pi G}\int\sqrt{-g} f(R)d^{4}x,
\end{equation}
where the Ricci scalar is replaced by some general function, $R\rightarrow f(R)$.  It was proved that such
a theory can be renormalized \cite{stele, Hindmarsh, tomb} from a perspective of a standard quantum field theory machinery.
Another success of $f(R)$ theory was recognized by Starobinsky \cite{starobinsky1, starobinsky2} who used it as a model of inflation in the early stage of the Universe leading to an effective cosmological constant. In the cosmology sector the
$f(R)$ gravity is found to be a very successful theory providing a natural explanation for dark energy, dark matter, cosmic bounce etc. without introducing some new unknown fields, exotic matter and other speculative notions \cite{Sotiriou:2008rp, cembranos1, nojiri1}, a brilliant hystorical review can be found in \cite{schmid}. The viability of the theory has also been discussed by several Solar system tests and constraints on the theory \cite{Guo1, Liu1} . One of the first drawbacks was the discovery of the Ostrogradsky instabilities and ghost degrees of freedom as the theory is based on the fourth order differential equation. Moreover, to get the unique solution of the Cauchy problem is extremely difficult \cite{Khodabakhshi:2018clk}. Recently, was discovered that with the Lagrange multiplier
constraint the theory was ghost free \cite{nojiri2} and in a nonlocal $f(R)$ gravity theory \cite{nojiri3} was found the same conclusion.
In metric $f(R)$ the following conditions must be satisfied so that the theory becomes free of Ostrogradsky instabilities \cite{wood} and ghost-free \cite{dolgov, sawicki}:
\begin{equation}
    \frac{df(R)}{dR}>0, \qquad \frac{d^{2}f(R)}{dR^{2}}\geq0.
\end{equation}
From our point of view we will treat the $f(R)$ theory as an effective toy theory of a quantum theory of gravity which has yet to be established. 
We will work in a so called metric formalism where the equations of motion are obtained by varying the action with respect to the metric. By doing so one obtains the following field equation\cite{Sotiriou:2008rp}
\begin{equation}
    f'(R)R_{\mu \nu} - \frac{1}{2}f(R)g_{\mu \nu} - (\nabla_{\mu} \nabla_{\nu} - g_{\mu\nu} \square)f'(R)=\kappa T_{\mu\nu},
 \label{fields}
\end{equation}
where $\kappa=8\pi G/c^{4}$ and as usual the stress-energy tensor is defined as
\begin{equation}
    T_{\mu\nu}\equiv\frac{-2}{\sqrt{-g}}\frac{\delta S_{m}}{\delta g^{\mu\nu}},
 \label{stress}
\end{equation}
where the prime denotes differentiation with respect to the argument, $\nabla_{\mu}$ is the covariant derivative and $\square \equiv \nabla_{\mu}\nabla^{\mu}$. Now turning to the cosmological setting we will use the FLRW metric (\ref{metrika}) with $k=0$, so that the resulting equations of motion are
\begin{equation}
    3\dot{H} + 3H^{2}=-\frac{1}{2f'}(\rho + 3p +f -f'R + 3Hf''\dot{R} + 3f''' \dot{R}^{2} + 3f''\ddot{R}),
    \label{friedmann1}
\end{equation}
\begin{equation}
    3H^{2}=\frac{1}{f'}\Big( \rho + \frac{1}{2}(Rf' - f)-3Hf''\dot{R}\Big),
    \label{friedmann2}
\end{equation}
where $H=\dot{a}/a$ is the Hubble parameter, the dot represents derivative with respect to time and $f'={\partial f}/{\partial R}$, $f''={\partial^{2}f}/{\partial R^{2}}$ and $f'''={\partial^{3}f}/{\partial R^{3}}$. 
\subsection{Cyclic solutions in $f(R)$ gravity} 
In order to study the 
dynamical properties of cyclic solutions in $f(R)$ gravity we should choose one of the related equations of motion to define a function $g(a,H,R)$ appearing in equation (\ref{kurva}). It seems natural to use the equation (\ref{friedmann2}), but the problem arises when one is
dividing the whole equation with $3Hf''$, to get $\dot{R}$ alone on the left side of the equation, as $H=0$ is actually a turnaround point we are here interested in. On the other hand, the equation (\ref{friedmann1}) does
not suffer from this feature, and simply by adding a new equation in
the dynamical analysis -- in order to take into account that the equation (\ref{friedmann1}) is now containing the second time derivative of the Ricci scalar, while adding a new degree of freedom $L$, we get the new set of equations
\begin{equation}
\dot{a}=a H   
\label{skala2}
\end{equation}
\begin{equation}
\dot{H}=\frac{1}{6}[R-12H^2] 
\end{equation}
\begin{equation}
\dot{R}=L,    
\end{equation}
\begin{equation}
    \dot{L}=\ddot{R}=h(a,H,R,L)
    \label{kurva2}
\end{equation}
where $L \equiv \dot{R}$ by definition and
\begin{equation}
    h(a,H,R,L)=\frac{1}{3f''}(6f'H^2 -\rho - 3p(\rho) -f - 3Hf''L - 3f''' L^{2}).
\end{equation}
We are now ready to perform the linear stability analysis; with the assumption of the fluid equation of state in the form $p(\rho)=w\rho$, by including matter ($w=0$) and radiation ($w=1/3$) component of the ideal fluid. The Jacobian $J$ then reads:
\begin{equation}
   J-\lambda \mathbbm{1}= \begin{pmatrix}
    H-\lambda       &    a    &   0  & 0 \\
    0       &    -4 H- \lambda    & \frac{1}{6} & 0 \\
    0      &  0   &     - \lambda & 1 \\
    \frac{\partial h(a,H,R,L)}{\partial a} & \frac{\partial h(a,H,R,L)}{\partial H} & \frac{\partial h(a,H,R,L)}{\partial R} & \frac{\partial h(a,H,R,L)}{\partial L} - \lambda
   \end{pmatrix} ,  
   \label{linearfr}
\end{equation}{}
we require $\det [J-\lambda \mathbbm{1}]=0$ elevated at the fixed point $a=a^{*}, H^{*}=0$, $R^{*}=0$, $L^{*}=0$, and this gives us the equation:
\begin{equation}
    \lambda^4 + \frac{f''(R) \left(6 a^4 \lambda ^2 f'(R)-3 a \rho_m-8 \rho_r\right)-6 \lambda ^2 f'''(R) \left(a^4 f(R)+a \rho_m+2 \rho_r \right)}{18 a^4 f''(R)^2}\Bigg|_{f.point}=0,
    \label{svojstvene1}
\end{equation}{}
with $f''(R)\neq 0$, the solution of the eigenvalue problem is:
\begin{equation}
    \lambda^2_{1,2} = \lambda_0^2 \pm \frac{ \sqrt{A+B^2 }}{36a^4 f''(0)^2}
\end{equation}{}
where:
\begin{equation}
    \lambda_0^2=\frac{1}{6} \left( \frac{2 \rho_r f'''(0)}{a^{*}{}^{4} f''(0)^2}+\frac{\rho_m f'''(0)}{a^*{}^3 f''(0)^2}+\frac{f(0) f'''(0)}{f''(0)^2}-\frac{f'(0)}{f''(0)}\right),
\end{equation}{}
\begin{equation}
    A=-72 a^*{}^4 f''(0)^2 \left(-3 a \rho_m f''(0)-8 \rho_r f''(0)\right),
\end{equation}{}
\begin{equation}
    B=-6 a^*{}^4 f(0) f'''(0)+6 a^*{}^4 f'(0) f''(0)-6 a \rho_m f'''(0)-12 \rho_r f'''(0).
\end{equation}{}
As discussed in Sec 2. the linear stability theory is inconclusive in determining the 
nature of fixed points, since centers
are not stable with respect to the effects of nonlinear corrections. Thus, the linear stability analysis needs to be 
further supported by reference to Claim 1 or Claim 2 discussed in Sec 2. in order to prove the existence of non-linear centers.
Then using Claim 2 it follows that the necessary condition $ \nabla I(a,H,R,L)=0$ around the fixed point $(a=a^*,H=0,R=0,L=0)$ 
in the case of dust and radiation in $f(R)$ gravity  leads to: 
\begin{equation}
    \frac{1}{2} (4 {a^{*}}+3) {a^{*}}{^2} f(0)+\frac{{\rho_{r}}}{a^{*}{^2}}-\rho_{m}=0.
    \label{zvijezda}
\end{equation}{}
Here we have used the fact that the conserved integral is in this case, of matter and radiation described by ideal fluid in $f(R)$ gravity, given by (see the discussion under Claim 2 in Sec.2):
\begin{equation}
I(a,H, R, L)=(3H^2f' - \frac{1}{2}(Rf' -f)+3Hf''L)(a^4 +a^3)- \rho_{m}a - \rho_{r}/a .    \end{equation}
In order to verify that the conserved integral at the fixed point indeed has an extremal nature, one should in principle also inspect the behaviour of $\nabla^2 I(a,H,R,L)$ at $a=a^*,H=0,R=0,L=0$. In the considered case  $\nabla^2I$ is at the fixed point equal to:
\begin{equation}
\small{\left(
\begin{array}{cccc}
 -\frac{2 \rho_r}{a^*{}^3}+ 3a^*f(0)(2a^*+1)   & 0 & 0 & 0 \\
 0 & 6 a^*{}^3 (a^*+1) f'(0) & 0 & 3 a^*{}^3 (a^*+1) f''(0) \\
 0 & 0 & -\frac{1}{2} a^*{}^3 (a^*+1) f''(0) & 0 \\
 0 & 3 a^*{}^3 (a^*+1) f''(0) & 0 & 0 \\
\end{array}
\right)}
\end{equation}{}

\subsubsection{Concrete numerical solutions of cyclic $f(R)$ model}
As an example one could choose a specific $f(R)$ and find the corresponding eigenvalues. In the following sections we will consider the polynomial form of the type:
\begin{equation}
    f(R)=a_1R+a_2R^2+a_3R^3+a_4.
    \label{carloni}
\end{equation}{}
Such polynomial form is both interesting because of its generality (since it represents the first terms in the Taylor expansion of any $f(R)$ function) and since such form of the curvature correction to Einstein-Hilbert action will typically arise when quantum loop corrections coming from the self-interaction of gravity are considered. 
Firstly, we will construct a numerical function of $f(R)$ which will lead to a cyclic Universe. By doing so we
will approximate this solution by a series similar to (\ref{carloni}) and perform a dynamical analysis to check
the viability and consistency of the two methods. The most simple example is to start with the scale factor as 
\begin{equation}
    a(t)=\frac{1}{2\Lambda}(1+c \sin (2\omega t)),
    \label{oscilating}
\end{equation}
where $\Lambda$, $c$ and $\omega$ are some real positive constants. If $c\in \langle -1,1 \rangle$ then 
the given scale factor corresponds to a well-defined and regular cyclic solution. On the other hand, if there is a time at which
$a(t_{min})=0$ then this scenario would lead to a singularity in $H$ and therefore in curvature,
$R \rightarrow \infty$. In this case the solution would not be strictly cyclic but it would lead to a so called Big
Crunch, i.e. the oscillation would be interrupted by the singular points reached in the configuration space $a$, $H$, $R$. Therefore, the required condition is: $a(t_{min})>0$ and for this reason the 
simpler expressions of the form $a(t)=(A\sin(\omega t)$, $A\cos(\omega t)$), commonly found in literature when discussing cyclic solutions, are not considered here. An earlier discussion of a reconstruction of such a solution, containing a singularity, can be found in \cite{carloni}. \\ \\
By expressing $a(R)$, the field equations (\ref{friedmann1}) and (\ref{friedmann2}) become a second
order differential equations in $f$ with respect to $R$.
The required functions expressed in terms of $R$ are
\begin{equation}
    a(R)=\frac{18 \Lambda  \omega ^2 \pm \sqrt{6} \sqrt{48 c^2 \Lambda ^2 \omega ^4+c^2 \Lambda ^2 R \omega ^2+6 \Lambda ^2 \omega ^4-\Lambda ^2 R \omega ^2}}{48 \Lambda ^2 \omega ^2+\Lambda ^2 R},
    \label{aodr}
\end{equation}
\begin{equation}
    \frac{da}{dt}=\frac{c \omega  \sqrt{1-\frac{(1-2 a(R) \lambda )^2}{c^2}}}{\lambda },
\end{equation}
\begin{equation}
    \frac{d^2a}{dt^2}=-\frac{2 \omega ^2 (2 a(R) \lambda -1)}{\lambda },
\end{equation}
\begin{equation}
    \frac{dR}{dt}=-\frac{12 c \omega ^3 \left(3 a(R) \lambda +c^2-1\right) \sqrt{1-\frac{(1-2 a(R) \lambda )^2}{c^2}}}{a(R)^3 \lambda ^3},
\end{equation}
and the field equation (\ref{friedmann2}) can be numerically solved to reconstruct a specific numerical $f(R)$ solving the equation:
\begin{equation}
  3\Big(\frac{\dot{a(R)}}{a(R)}\Big)^{2}=\frac{1}{f'}\Big( \frac{\rho_0}{a(R)^3} + \frac{1}{2}(Rf' - f)-3\frac{\dot{a(R)}}{a(R)}f''\dot{R}\Big).
  \label{numerical}
\end{equation}
By choosing $\rho_0=1$, $c=0.8$, $\omega=2$ and $\Lambda=0.01$ with the initial conditions:
\begin{equation}
    f(0)=a_4=1.5\cdot 10^{-6}, \qquad f'(0)=a_1=-10^{-8},
    \label{initial}
\end{equation}
we obtain the solution which is depicted in Fig 2. By using the following values
\begin{equation}
    a_1=-10^{-8}, \qquad a_2=-2.5\cdot 10^{-10}, \qquad a_3=-1.7 \cdot 10^{-13}, \qquad a_4=1.5\cdot 10^{-6},
    \label{param}
\end{equation}
 in expression (\ref{carloni}), it is obvious that the numerical $f(R)$ can be approximated with (\ref{carloni}) within the error of magnitude of the  order of $10^{-5}$. For larger values of $R$ the error is increasing, which does not come as a surprise -- since in the high curvature regimes the higher orders of $R$, determining the further features of the full numerical solution not contained in the approximated solution, will become important and need to be included to effectively mimic the quantum effects of gravity.
\begin{figure}[h]
\centering
    \includegraphics[width=0.9\linewidth]{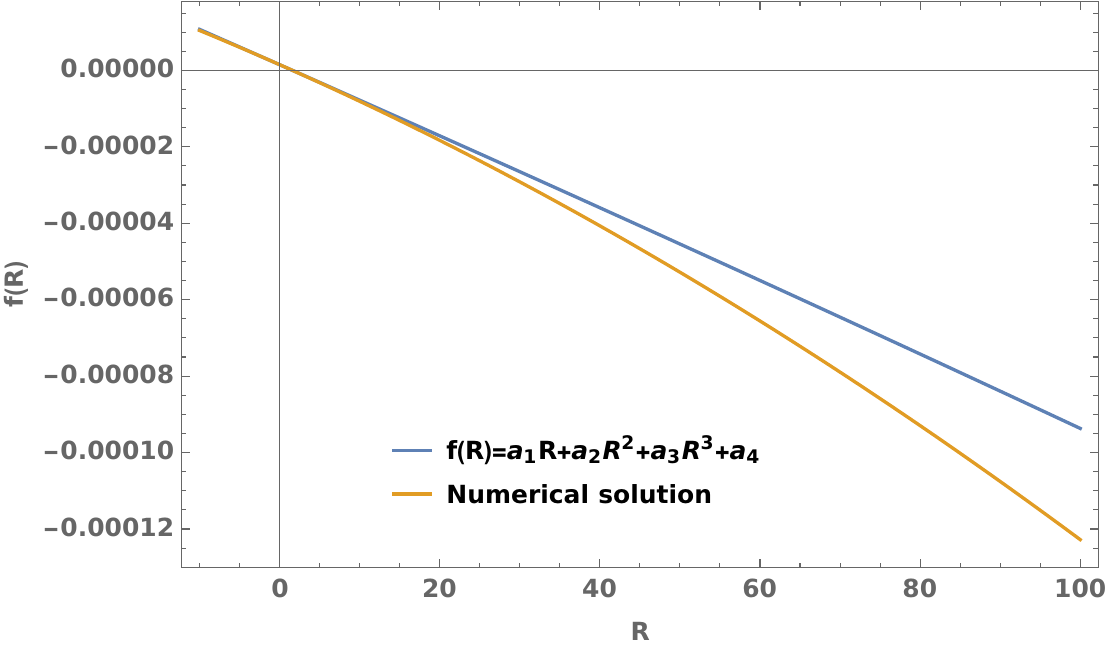}
       \caption{The numerical solution of eq. (\ref{numerical})   
       with the scale factor evolution given by (\ref{oscilating}), and the initial conditions given by  (\ref{initial}) (orange line) compared to analytical approximation (\ref{carloni}) with the parameters given by (\ref{param}) (blue line)}
\end{figure}
\subsubsection{Dynamical analysis of $f(R)$ third order polynomial model}
\begin{figure}[h]
\centering
    \includegraphics[width=0.9\linewidth]{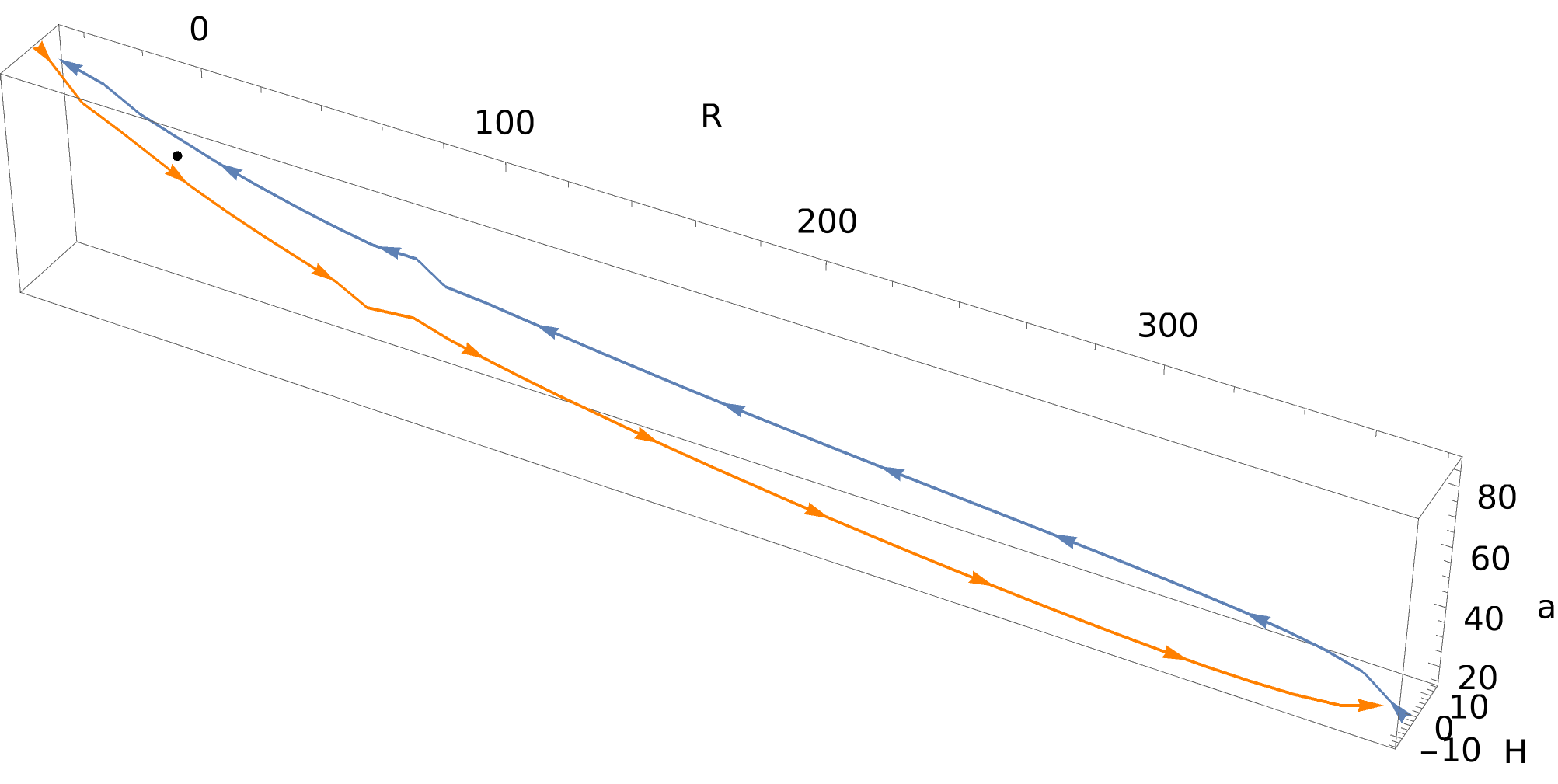}
       \caption{Phase portrait (a,R,H) of the numerical solution of (\ref{numerical}) with the values $\Lambda=0.01$, $\omega=1$, $c=0.8$ and $\rho_0=1$. The dot corresponds to the center point $a^*=69$, $H=0$, $R=0$. The orange and blue line represent two different branches as $H=\dot{a}/a$ is quadratic in (\ref{numerical}), they collide at the maximum and the minimum value of the scale factor, $a$, taking place at $H=0$. 
       } \label{rekonstrukcija}
\end{figure}

Now we are ready to analyse the cyclic solutions in the $f(R)$ gravity given by (\ref{carloni}), and we expect that
with the same parameters the dynamical analysis should be consistent with the existence of cyclic solutions. Considering the same $f(R)$ as discussed earlier,
\begin{equation}
    f(R)=a_1R+a_2R^2+a_3R^3+a_4,
\end{equation}
with
\begin{equation}
    a_1=-10^{-8}, \qquad a_2=-2.5\cdot 10^{-10}, \qquad a_3=-1.7 \cdot 10^{-13}, \qquad a_4=1.5\cdot 10^{-6},
\end{equation}{}
for the choice of parameters $w=0$, $\rho_r=0$ and $\rho_m=1$, one can calculate $a^{*}$ from the necessary condition for the existence of a nonlinear center (\ref{zvijezda}) at the fixed point $(a=a^{*},H=0,R=0,L=0)$, using the requirement that at this fixed point the conserved integral has an extremal value.  This yields 
\begin{multline}
 a^*=  \frac{1}{4} \Bigg(\frac{f(0)}{\sqrt[3]{4 \sqrt{2} \sqrt{8 f(0)^4 \rho_{m}^2-f(0)^5 \rho_m}+16 f(0)^2 \rho_{m}-f(0)^3}}+ \\
 \frac{\sqrt[3]{4 \sqrt{2} \sqrt{8 f(0)^4 \rho_{m}^2-f(0)^5 \rho_{m}}+16 f(0)^2 \rho_{m}-f(0)^3}}{f(0)}-1\Bigg) \simeq 69.09,
\end{multline}
By comparing the value of scale factor at the resulting extremal point of $I(a,H,R)$ with the value of scale factor at the fixed point, given by analytical approximation (\ref{aodr})
\begin{equation}
    a(R=0)=\frac{\sqrt{\left(8 c^2+1\right) \lambda ^2 \omega ^4}+3 \lambda  \omega ^2}{8 \lambda ^2 \omega ^2}\simeq 68.4,
\end{equation}
one can conclude that the two fixed points are in an excellent agreement given the involving approximation. Consequently, the numerical $f(R)$ can be effectively modeled as a shift from the (\ref{carloni}):
\begin{equation}
    f(R)_{numerical}=f(R)+\Delta f(R),
\end{equation}
which corresponds to a shift in the $a^*(R=0)\rightarrow a^* + \Delta a^*$. In order to inspect the cyclic solution within the linear analysis we need to find the eigenvalues from (\ref{svojstvene1}) for the specific $f(R)$ model. Using the same values as in the numerical procedure we get that the eigenvalues are:
 $\lambda_{1,2}\approx 25.4i$ and $\lambda_{3,4}\approx 1.29i$. We see that in the considered example such type of the fixed point, containing imaginary eigenvalues and being stable by the virtue of the extremal nature of $I(a,H,R)$ at this point, leads to a cyclic solution.
  Similarly to Fig \ref{Bfunr}. the phase portrait of this solution is given in Fig \ref{rekonstrukcija}.

\section{Cyclic cosmologies with dynamic dark energy}
\subsection{Introduction} 
The usual assumption, invoked in the standard cosmological model to solve the contradiction between the observed accelerated expansion of the Universe and the attractive nature of gravity in general relativity, is to introduce a small constant term (the cosmological constant) into the field equations. The standard interpretation of this constant is that it represents the vacuum energy contribution, an interpretation which -- as it is well known -- opens new severe problems due to its small observed value compared to the huge value predicted by the quantum field theory order of magnitude estimate
\cite{wein1,rev}. At the same time, the standard cosmological model gives a currently satisfactory fit to the empirical data related to the cosmological evolution, which is -- together with its simplicity -- the main reason for its popularity. However, introducing a constant cosmological term represents only one among several other related possibilities. Specifically, there are no reasons against a much more general option -- that the cosmological term is not a constant but a dynamic quantity. Moreover, such an option can be motivated by additional theoretical considerations. If cosmological term is understood as coming directly from the vacuum energy density, then quantum field theory considerations on curved spacetime can motivate the running vacuum models, in which its energy becomes dynamic \cite{vacuum1, vacuum2}. On the other hand, if the cosmological term is understood as a lower curvature effective contribution coming from some new theory of (quantum) gravity then it can also be expected that this contribution would in general be dependent on the considered energy regime. In connection with this reasoning, it is worth to note that a running nature of the couplings of the theory is something that comes as a usual consequence of effective field theories, as it is for instance also discussed in the asymptotically safe gravity approaches \cite{safe}. All these reasons speak strongly in favour of the need to analyse the cosmological consequences and models in the framework of dynamic cosmological term, which was recently discussed in various works. These works demonstrated that some of dynamic energy models are in a very good agreement with the empirical data \cite{dde1,dde2,dde3,dde4}.
Basing ourselves on this motivation, we want to apply our general discussion of cyclic cosmologies to the models of dynamic dark energy and for the first time present some models of cyclic cosmology in this framework. \\ \\
The cosmological term considered as an explicit function of time, 
$\Lambda(t)$, in general gives rise to a non-autonomous set of differential equations, which cannot be analysed using the discussed techniques, and is therefore outside the scope of this work. The assumption we take in this section is that the dynamics of the cosmological term can be expressed as a dependence on the scale factor, curvature and Hubble parameter, $\Lambda(a, H, R)$. Under this assumption 
the cosmology with dynamic dark energy can be put in the form given 
by the equations  (\ref{skala})-(\ref{kurva}), with
\begin{equation}
g(a,H,R)=
\frac{1}{\frac{\partial \Lambda}{\partial R}}[\frac{1}{3}(R-12H^2)H - \frac{1}{3}\frac{d\rho(a)}{da}H -\frac{\partial \Lambda}{\partial a}H - \frac{1}{6}\frac{\partial \Lambda}{\partial H}(R-12H^2)]. 
\label{dindark}
\end{equation}
Now we can simply apply the theorem for 
the existence of non-linear centers on such theory where cosmological term is given by 
$\Lambda(a,H, R)$. \\ \\ Claim 3: \textbf{If dynamic dark energy is a function with the following property: {$\mathbf{\Lambda(a,H, R)=\Lambda(a, -H, R)}$}, and the system of equations describing the cosmological evolution in such theory of dynamic dark energy has a center at some value  $\mathbf{a=a^{*}}$ with $\mathbf{H^{*}=0}$ and $\mathbf{R^{*}=0}$, determined by the linear stability theory from the stability matrix (\ref{matrica}) -- then there exists a non-vanishing neighbourhood around this point such that all trajectories inside of it will correspond to symmetric cyclic cosmological solutions.}
Proof: If $\Lambda(a,H, R)=\Lambda(a, -H, R)$ and its corresponding function, $g(a,H,R)$, is determined by the equation (\ref{dindark}), then the system of equations describing such cosmological evolution is a reversible dynamic system. Then, according to the presented Claim 1, sufficiently close to the linear center of this system of equations all trajectories will correspond to cyclic cosmological solutions. 
\subsection{$\Lambda(a)$ dynamic dark energy model}
We will now consider a specially convenient form of simple dynamic dark energy models where the cosmological term -- absorbing all contributions modifying the standard Friedmann equations whatever be their cause (for instance, modified gravity or new types of cosmological fluid) -- can be expressed as a function only of the scale factor, $\Lambda(a)$. Furthermore, here and in the following sections we will assume that dynamic dark energy can be treated as non-interacting with matter fields, so that the evolution of energy density and pressure is still having the standard form. It then follows that in this case 
the problem can be reduced to a one-dimensional system of the following form:
\begin{equation}
\frac{a'(t)}{H_{0}}=\pm \sqrt{\Omega^{rad}_{0}a^{-2}+\Omega^{mat}_{0}a^{-1}+ \Lambda(a)a^2} . 
\label{lambdaa},
\end{equation}
where for the convenience we introduced the radiation and 
matter densities today, $\Omega_{0}^{rad}$ and $\Omega_{0}^{mat}$, as
well as the dark energy assumed to be expressed with respect to the critical density, viz. $\Lambda(a) \equiv \rho_{\Lambda}/(3H_{0}^2/8 \pi G)$. \\ Although this is a first order autonomous system, the existence of cyclical solutions is possible by virtue of existence of two branches of solutions, corresponding to expansion (positive branch) and contraction (negative branch). In order to enable both the transition from the contracting to expanding phase (the cosmological bounce) and the vice versa (the cosmological turnaround) these branches need to connect at two different fixed points of the equation (\ref{lambdaa}). These fixed points in the case of cyclic solutions correspond to the minimal and maximal values of the scale factor, $a_{min}$ and $a_{max}$. For this reason the solutions of the equation (\ref{lambdaa}) need to be constrained to the region 
$a_{min} \leq a \leq a_{max}$ and outside this region it follows:
\begin{equation}
\Omega^{rad}_{0}a^{-2}+\Omega^{mat}_{0}a^{-1}+ \Lambda(a)a^2<0 .
\end{equation}

Regarding the stability of solutions in the case of the positive branch, corresponding to expansion, the points in the nearby region of $a_{min}$ need to be repelled from the fixed point, while they need to be attracted towards it in the case of the negative branch, which is corresponding to contracting phase of the Universe. The reverse if true for the second fixed point at $a_{max}$. In this way both fixed points need to act as semi-stable fixed points enabling the transition from the contraction to expansion and vice versa. Therefore, considering the linear stability for the positive branch it follows
\begin{equation}
2a_{min} \Lambda(a_{min}) + a_{min}^{2}\frac{d \Lambda(a)}{da}|_{a=a_{min}}-2a_{min}^{-3} \Omega^{rad}_{0}- a_{min}^{-2} \Omega^{mat}_{0}>0
\end{equation}
and
\begin{equation}
2a_{max} \Lambda(a_{max}) + a_{max}^{2}\frac{d \Lambda(a)}{da}|_{a=a_{max}}-2a_{max}^{-3} \Omega^{rad}_{0}- a_{max}^{-2} \Omega^{mat}_{0}<0,
\end{equation}
while the opposite inequalities need to hold for the negative branch of equation (\ref{lambdaa}). 
\\ \\
As a very simple example of cyclic cosmology in $\Lambda(a)$ let us consider the following function
\begin{equation}
\Lambda_{trig}(a)= \frac{1-(k-a)^{2}}{a}- (\Omega^{mat}_{0} + \Omega^{rad}_{0}),
\label{cos}
\end{equation}
where $k$ is a constant. 
In this case the modified  Friedmann equation (\ref{lambdaa}) simply reduces to $\dot{a}=f(a) = \pm \sqrt{(1-(k-a)^2)a}$, and has two fixed points corresponding to $a_{min}$ and $a_{max}$ given by:
$a_{min,max}=k \mp 1$ at which the positive and negative branch meet. Demanding that the scale factor 
always stays positive, we have clearly $k>1$ and it is straightforward to check that under this condition
$df/da>0$ at $a_{min}$ for the positive branch of the solution, and $df/da <0$ at $a_{min}$ for the negative branch of the solution -- i.e. the points on the positive branch will be repelled from this fixed point, while the points on the negative branch will be attracted towards it, enabling the transition of the Universe from the contracting into the expanding phase. Conversely, at $a_{max}$ it simply follows that  $df/da <0$ for the positive branch and $df/da>0$ for the negative branch -- so that the late time expansion of the Universe changes into a contracting phase leading to a new cosmological bounce. The model (\ref{cos}) can be criticised due to the fact that the energy densities enter into the functional dependence of $\Lambda(a)$, which may be viewed as not natural. However, this could be understood simply in terms of modeling a situation in which, in a given regime, the contributions of matter and radiation densities are compensated by the opposite contribution of the dynamic dark energy. As we will discuss in the following, the matter and radiation can in a more general scenario be introduced as a perturbation around this solution. \\ \\
In fact, the model described by (\ref{cos}) leads to the analytical oscillating solutions for the scale factor given by: $a(t)=k -cos(t)$, and its discussed dynamic properties therefore do not come as a surprise. In order to study more general and realistic scenarios we can consider adding arbitrary correction terms containing powers of the scale factor to model (\ref{cos}): $\Lambda(a)= \Lambda_{trig}(a) + \sum_{n}c_{n} a^{n}$. The considerations of dynamic properties of such $\Lambda(a)$ model, following the general discussion given earlier, will then lead to constraints on the coefficients $c_{n}$ in order to lead to cyclic cosmologies. Considering the corrections to the first order, the
fixed points of the equation (\ref{lambdaa}) will then be given by the solution of the 
associated third order algebraic equation. In order to have cyclic cosmologies there need to exist two real and positive solutions of this equation, corresponding to $a_{min}$ and $a_{max}$. Since one of the solutions, $a=0$, can be discarded as the physical solution of interest, we are left with the following two solutions: 
\begin{equation}
a_{min, max}= \frac{-(c_{0}+2k) \pm \sqrt{(c_{0}+2k)^{2}-4(c_{1}-1)(1-k^{2})}}{2(c_{1}-1)}, 
\end{equation}
where it needs to be demanded that both solutions are real and positive, which constraints the values of 
parameters $c_{0}$ and $c_{1}$ . 
\\ \\
There is also a different and more general class of $\Lambda(a)$ models we can construct in order to obtain cyclic cosmologies which also approach the $\Lambda$CDM model in the period between the bounce and turnaround. 
To fulfill the discussed conditions for cyclic evolution the value of dynamic dark energy needs to become negative while approaching both the bounce and turnaround fixed points, at $a_{min}$ and $a_{max}$ respectively, while to reproduce the $\Lambda$CDM evolution it needs to approach approximately constant and positive values during the radiation, matter and dark energy dominated phase. One possible class of such functions is given by:
\begin{equation}
\Lambda(a)= \Lambda_{0}(1- \frac{g(a)}{a^{k}} - h(a) a^{m}),    
\end{equation}
where $m>0$, while the function $h(a)$ needs to satisfy $h(a) a^{m} \approx 0$ for $a \ll a_{max}$ and be consistent with the existence of a fixed point at $a_{max}$, while $k\geq 4$ and the function 
$g(a)$ needs to satisfy $g(a)/a^k \approx 0$ for $a_{min} \ll a$ and 
it moreover needs to be consistent with the existence of a fixed point at $a_{min}$. We plot the phase portraits of one example of this class of models in Fig. 3., together with examples for other types of models discussed in this section . 
\\ 
\begin{figure}[h]
\centering
    \includegraphics[width=0.8\linewidth]{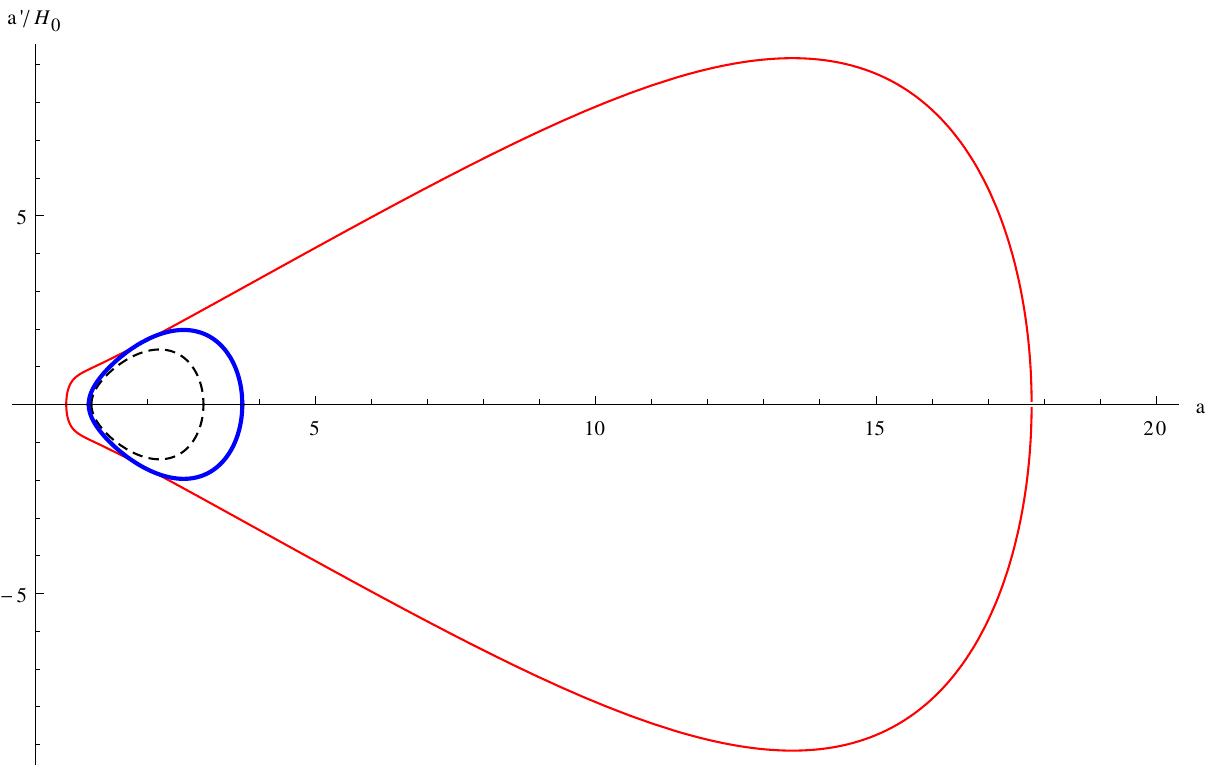}
       \caption{Various models of $\Lambda(a)$ dynamic dark energy leading to cyclic cosmologies:
       $\Lambda_{trig}(a)= \frac{1-(c-a)^{2}}{a}- (\Omega^{mat}_{0} + \Omega^{rad}_{0})$ (black dotted line), $\lambda(a)=\Lambda_{trig}(a)+ c_{0}a + c_{1}a^{2}$ (blue thick line), $\Lambda(a)= \Lambda_{0}(1- \frac{g(a)}{a^{k}} - h(a) a^{m})$ (red full line). Here the parameters are chosen to be: $c=2$, $\Omega^{mat}_{0}=0.3$, $\Omega^{rad}_{0}=10^{-5}$, $\Lambda_{0}=0.69$, $c_{0}=0.1$, $c_{2}=0.01$, $g(a)=0.1$, $k=4$, $m=4$, $h(a)=10^{-5}$.} \label{lamoda}
\end{figure}

\subsection{$\Lambda(R)$ model}
On the other hand, one could also inspect the specific case where
dark energy is a function only of the Ricci scalar, namely $\Lambda(R)$.
In this approach the action of the theory can be simply thought as a specific example of a $f(R)$ theory:
\begin{equation}
f(R)=-2\Lambda(R) + F(R).
\end{equation}{}
Therefore the techniques presented in Sec.3 can be applied in the setting of $\Lambda(R)$ dark energy. As an example we will refer to the article \cite{cik12} where a specific $\Lambda(R)$ was discussed in connection with  cyclic cosmological solutions. Here we will show that using a different approach, namely a dynamical analysis, the same conclusion follows. The specific $f(R)$ chosen in \cite{cik12} is given by
\begin{equation}
   f(R)= -2 \Lambda(R)+ R+a_2 R^2+a_3 R^3,
\end{equation}
where:
\begin{equation}
a_2=\frac{3 \Lambda_0 +36 \Lambda_0  R_2-2}{1-12 R_2} \qquad a_3=\frac{-2 \Lambda_0 -12 \Lambda_0  R_2+1}{1-12 R_2},
\end{equation}{}
and $\Lambda(R)$ is given by
\begin{equation}
\Lambda(R)=\Lambda_0-\frac{A \left(1-\frac{R-R_{min}}{B}\right)}{\frac{(R-R_{min})^2}{B^2}+1}.
\end{equation}{}
From equation (\ref{svojstvene1}) without matter fields $\rho_{r,m}=0$ the eigenvalues are
\begin{equation}
    \lambda_{1,2}=0, \qquad \lambda_{2,3}=\pm \frac{i \sqrt{f'(R) f''(R)-f(R) f'''(R)}}{\sqrt{3} f''(R)},
\end{equation}{}
it is easy to see that the solutions leading to a cyclic Universe must satisfy the condition
\begin{equation}
    f(R) f'''(R)>f'(R) f''(R). 
\end{equation}{}
Using the same parameters as in \cite{cik12}
$A=0.5$, $R_{min}=-0.5$, $B=-0.1$, $\Lambda_0=0.0005$ and $R_2=-2$ the resulting eigenvalues equals to $\lambda\approx\pm 0.582i$.
Evidently the linear stability test is not conclusive due to the fact that two eigenvalues are equal to zero.
\\ \\
The conserved quantity in $\Lambda(R)$ model can be expressed as:
\begin{equation}
I(H, R, L)=3H^2f' - \frac{1}{2}(Rf' -f)+3Hf''L.    
\end{equation}
The conserved quantity $I(H, R, L)$ must satisfy:
\begin{equation}
\nabla I(H, R, L)=\left(
\begin{array}{c}
 3 L f''(R)+6 H f'(R) \\
 3 H L f^{(3)}(R)+\left(3 H^2-\frac{R}{2}\right) f''(R) \\
 3 H f''(R) \\
\end{array}
\right)=0,
\end{equation}
which is evidently satisfied at $H=0$, $R=0$ and $L=0$.

\section{Cyclic cosmologies in modified torsion based $f(T)$  theories of gravity}
\subsection{A short review of $f(T)$ 
gravity}
It is a well known concept to introduce torsion in a theory of gravity. In fact, one can construct a theory of gravity equivalent to General Relativity (GR) by replacing the description of spacetime which is curved with spacetime that has a non-vanishing torsion. The equivalent theory is called Teleparalel Equivalent to General Relativity (TEGR). In contrast to GR, where the curvature scalar, $R$, is contained in the Einstein-Hilbert action, in TEGR the curvature scalar is replaced with the so called torsion scalar, $T$. The action in TEGR is given by \cite{aldro}:
\begin{equation}
    \mathcal{S}_{TEGR}= \frac{c^{4}}{16 \pi G}\int h T d^{4}x,
    \label{akcijategr}
\end{equation}
where $h$ is the determinant of the tetrad field $h^{a}{}_{\mu}$, which is the dynamical degree of freedom and is related to the metric tensor (metric compatibility condition) $h^{a}{}_{\mu}h^{b}{}_{\nu}g^{\mu \nu}=\eta^{ab}$, where $\eta^{ab}$ is the Minkowski metric with the  signature $\eta^{ab}=$diag$(1,-1,-1,-1)$. The torsion scalar, $T$, is a rather special scalar constructed in such a way to lead to a relationship with $R$ in a way that they differ only by a total divergence:
\begin{equation}
    R=-T - 2 \nabla^{\mu}T^{\nu}_{\mu \nu}.
\end{equation}{}
where $T^{\rho}_{\mu \nu}$ is the torsion tensor defined as an antisymmetric part of an arbitrary affine connection: $\Gamma^{\rho}_{\nu \mu} - \Gamma^{\rho}_{\mu \nu}$. The torsion scalar is obviously always zero if $\Gamma^{\rho}_{\mu \nu}$ is a standard curvature Levi-Civita connection. From this construction it is easy to see that the resulting equations of motion from action (\ref{akcijategr}) are equivalent to those of GR. Good reviews of TEGR can be found in \cite{cai, yang, Myrzakulov, awad}.
In the same spirit as in $f(R)$ theories, one could replace the torsion scalar with an arbitrary function $f(T)$ in the action integral. The resulting class of theories is then called the $f(T)$ theories of gravity. The $f(T)$ theories are drastically different from those of $f(R)$ gravity, with one of the most striking problem given by the apparent break of local Lorentz invariance \cite{barrow1, barrow2}. The recent development of the theory found some interesting aspects of the structure of theory \cite{li2, bajerano, blagojevic, nester, krssak, golovnev, hohmann2} but the problem is still not well understood. Nevertheless, the $f(T)$ theories are very successful in the cosmological setting, as dark energy models and inflationary models, in strong gravity regimes, and are able to ensure the consistency with the solar system test \cite{ferugga, iorio, qi}. For this reason the cosmological behaviour of solutions is worth investigating in the context of the bouncing and cyclic cosmology \cite{cai3}. One of the main difference with respect to $f(R)$ is that remarkably the equations of motions are of the second order only, the same order as in GR, which could simplify considerably the dynamical system analysis. Recently, a "no go" theorem appeared in \cite{hohmann} where it was claimed that it leads to impossibility of a cyclic Universe in $f(T)$ gravity. We will inspect this problem further in the following section. 
\\ \\
The action in $f(T)$ gravity is
\begin{equation}
    \mathcal{S}= \frac{c^{4}}{16 \pi G}\int h f(T) d^{4}x.
\end{equation}{}
By varying the action with respect to the tetrad field the field equations of motion are obtained \cite{krssak}:
\begin{multline}
 h^{-1}f_{T}\partial_{\nu}(hS_{a}^{\;\; \mu \nu}) + f_{TT}S_{a}^{\;\; \mu \nu}\partial_{\nu}T  \\
 -f_{T}T^{b}_{\;\; \nu a}S_{b}^{\;\; \nu \mu} +\frac{1}{4}f(T)h_{a}^{\;\; \mu}
 +f_{T}A^{b}_{\;\; a \nu}S_{b}^{\;\; \nu \mu}=\kappa \Theta_{a}^{\;\; \mu},
\end{multline}
where $f_T=df(T)/dT$ and $f_{TT}=d^2f(T)/dT^2$, the torsion tensor is defined as the antisymmetric part of an arbitrary connection  $\Gamma^{\rho}_{\;\; \mu \nu}$
\begin{equation}
    T^{\rho}_{\;\; \mu\nu}= \Gamma^{\rho}_{\;\; \nu \mu} -  \Gamma^{\rho}_{\;\; \mu \nu},
\end{equation}
the superpotential is
\begin{equation}
S^{\rho \mu \nu}= K^{\mu\nu\rho} - g^{\rho \nu}T^{\lambda \mu}_{\lambda} + g^{\rho\mu }T^{\lambda \nu}_{\lambda}  ,
\end{equation}{}
with the contorsion tensor defined as
\begin{equation}
    K^{\rho}_{\;\; \mu \nu}=\frac{1}{2}(T_{\mu \;\; \nu}^{\;\;\rho}  + T_{\nu \;\; \mu}^{\;\;\rho}      -   T^{\rho}_{\;\; \mu\nu}  ),
\end{equation}{}
the spin connection is
\begin{equation}
    A^{a}_{\;\; b \mu}=h^{ a}_{\;\; \nu} \partial_{\mu}h^{\;\;\nu}_{ b}+h^{ a}_{\;\; \nu} \Gamma^{\nu}_{\;\; \rho \mu}h^{\;\;\rho}_{ b}\equiv h^{ a}_{\;\; \nu} \nabla_{\mu} h^{\;\;\nu}_{ b},
\end{equation}{}
and finally the torsion scalar is given by
\begin{equation}
    T=T^{a}_{\;\; \mu\nu}S_{a}^{\;\; \mu\nu}.
\end{equation}{}
The stress energy tensor is commonly 
\begin{equation}
   \Theta _{a}^{\;\;\rho}=\frac{-1}{h}\frac{\delta (h\mathcal{L}_{matter})}{\delta h^{a}_{\;\;\rho}}.
\end{equation}
In the cosmological setting we will use the tetrad field metric compatible with the FLRW spacetime
\[h^{a}_{\;\;\mu} = 
 \begin{pmatrix}
  1 & 0 & 0 & 0 \\
  0 & a(t) &  &  \\
  0  & 0  & a(t) r & 0  \\
  0 &  0 &  &  a(t) r \sin \theta
 \end{pmatrix}
 \]
 for this tetrad choice the corresponding spin connection is \cite{krssak}
 \begin{equation}
     A^{1}_{\;\; 2 \theta}=-1, \qquad A^{1}_{\;\; 3 \phi}=-\sin \theta, \qquad A^{2}_{\;\; 3 \phi}=-\cos \theta.
 \end{equation}
 With $T=-12 H^2$, the Friedmann equations of motion in $f(T)$ gravity turn out to be \cite{krssak}
 \begin{equation}
     6H^2f_T+\frac{f(T)}{4}=4\pi \rho,
     \label{fritele1}
 \end{equation}{}
 \begin{equation}
     2\dot{H}(24 H^2 f_{TT}-f_T)=4\pi (\rho + p).
     \label{fritele2}
 \end{equation}{}
 From the perspective of a dynamical analysis, the equations of motion in the covariant $f(T)$ gravity are much simpler than
 those of $f(R)$ gravity, as they are of the second order in contrast to the fourth order in $f(R)$. Therefore it is not a
 surprise that the $f(T)$ gravity is frequently explored as a dynamical system \cite{awad, bahamonde}.
 \subsection{On the existence of cyclic solutions in $f(T)$ cosmologies}
 An extensive dynamical analysis of $f(T)$ cosmologies was earlier conducted in \cite{hohmann} where it was concluded that cyclic solutions are prohibited in $f(T)$ gravity since the conditions for a bounce and turnaround are mutually contradictory and can thus not be at the same time realised for a single $f(T)$ function. In the present study we have reached the opposite conclusion: that cyclic cosmological solutions are actually 
 possible in $f(T)$ theories of gravity. The first type of proof for this claim is simply given by constructing the specific counter-examples. Namely, in order to prove that cyclic solutions are possible in $f(T)$ gravity, assuming the stress-energy tensor to be given by dust and radiation, it is sufficient to find a $f(T)$ function which leads to cyclic solutions in $a(t)$ and $H(t)$ when equations (\ref{fritele1}) and (\ref{fritele2}) are solved. We reconstruct such functions in the following subsection 5.4 and thus demonstrate the existence of cyclic solutions. Furthermore, by carefully inspecting the arguments used against the existence of cyclic cosmologies in $f(T)$ gravity elaborated in \cite{hohmann} we will try to show that the earlier conclusion on impossibility of cyclic solutions was not justified. \\ \\
 The earlier conclusion on the impossibility of cyclic solutions was derived as a consequence of the statement 3 presented in \cite{hohmann}. If we introduce $W=f(T) -T + 6H^{2} + 12H^{2}(df(T)/dT -1)$ and $W_{H}=dW/dH$, $W_{HH}=dW_{H}/dH$, then this statement reads \cite{hohmann}: At $H=0$ we have $H'(t)  \neq 0$ if and only if $W>0$, $W_{H} = 0$ and $W_{HH} \neq 0$, where
 \begin{itemize}
     \item For $W_{HH}<0$ we have $H'(t)>0$ and thus a bounce,
     \item while for $W_{HH} >0$ we have $H'(t)<0$ and thus a turnaround. 
     \end{itemize}
     This statement can be directly derived if one inspects the modified Friedmann equation in $f(T)$ gravity.
It was then concluded that since these two conditions are mutually exclusive, they can not be simultaneously realized in any $f(T)$ theory. It is at this specific point of the argument that the problem in the conclusion arises. It is true that the above stated conditions, $W_{HH}<0$ and $W_{HH} >0$ are mutually exclusive, but these two conditions are not realized at the same point of the cosmological time, but at different time points during the cyclic cosmological evolution. The difficulty in seeing this clearly comes from the fact that $H$ is by definition not an uniquely determined parameter in cyclic cosmology. The value $H=0$ corresponds both to bounce and turnaround, while -- if the cyclic solutions are symmetric with respect to the bounce -- the values of $H^{2}$ will be equal for the symmetric points in the phases of contraction and expansion. Since in the FLRW spacetime the torsion, which is a fundamental dynamical degree of freedom, is given by $T=-12H^{2}$, this double-valued nature of $H$ will also be reflected in the structure of field equations and properties of solutions. Due to this, the nature of dependence between $a$ and $H$, will be double-valued, as will be for instance seen in equation (84). For this reason, there will actually exist two different $f(T)$ functions which are leading to the same oscillatory $a(t)$ solutions, as will be demonstrated in section 5.4 by a reconstruction. Therefore, if one of these $f(T)$ functions is used in the Friedmann equation, and the problem is analysed from the point of view of change of the scale factor in time, the complete cyclic solution will be obtained. However, if one wants to analyse the problem from the point of view of $a-H$ dependence, for instance employing $W(H)$ as in the presented claim 3 in \cite{hohmann}, then the whole range of this dependence needs to be taken into account, and one needs to consider both branches of solutions, corresponding to both branches of $f(T)$, to cover the full phase plane. As those functions lead to the same $a(t)$ when the Friedmann equation is solved, they are in this sense dynamically indistinguishable. On the other hand, from the point of view of $a-H$ phase plane, one branch will fulfill the conditions for a bounce and other for a turnaround and, due to the double-valued nature of this dependence, they both need to be taken into account. We will try to demonstrate this properties of solutions in detail in section 5.5, and they are also clearly visible in Figure 6.

 \subsection{Vacuum cosmology}
 A simple example of $f(T)$ cosmology can be the case where the matter field energy-density contributions are negligible with respect to 
 torsion. As a result the matter fields are zero in the field equations in this regime:
 \begin{equation}
     6H^2f_T+\frac{f(T)}{4}=0,
     \label{fritele1vak}
 \end{equation}{}
 \begin{equation}
     2\dot{H}(24 H^2 f_{TT}-f_T)=0.
     \label{fritele2vak}
 \end{equation}{}
 By analysing the first equation of this system (\ref{fritele1vak}) one can obtain a remarkable result that with a specific $f(T)$ all scale factor functions, $a(t)$ -- and thus all kind of possible oscillatory functions -- are the solutions of the field equations. Namely, the equation can be written as:
 \begin{equation}
     \frac{f(T)}{4}-\frac{1}{2} T f'(T)=0,
 \end{equation}{}
 and can be thought as a first order differential equation in $f(T)$, whose solution is
 \begin{equation}
     f(T)=C \sqrt{-T}.
 \end{equation}{}
 It is straightforward to show that this specific $f(T)=C \sqrt{-T}$ is also a solution to the second field equation (\ref{fritele2vak}). As a consequence of this, for an arbitrary $a(t)$ the field equations are automatically solved -- i.e. the contribution of $\sqrt{-T}$ in the $f(T)$ lagrangian does not change the equations of motion that result from the Friedmann equations. Therefore, we can conclude that the $f(T)$ lagrangians with the transformation of the type:
 \begin{equation}
     f(T) \rightarrow f(T)+C\sqrt{-T},
 \end{equation}{}
 are all symmetric and the equations of motion remain the same under this transformation when the FRLW geometry is assumed. Then a question of potential generalisation arises: ,,Does such kind of transformation, for which all functions describing the spacetime geometry are satisfying the gravitational field equations, also exists for other spacetime geometries apart from FRWL spacetime?'' If this is true than all this contributions cannot by themselves be a viable $f(T)$ modification. This is a strong restriction of $f(T)$  gravity as only viable $f(T)$ are those excluded from all this symmetric contribution $f(T)\rightarrow f(T) + f(T)_{symm. contr.}$, $f(T) \ne f(T)_{symm. contr.}$. Further investigation on this problem is necessary which is beyond the scope of this paper.
 
 \subsection{Matter and radiation era}
 
 When the matter contribution is included then the field equations have the full form given by (\ref{fritele1}) and (\ref{fritele2}). From a simple relation between the torsion scalar and the Hubble parameter, $T=-12H^2$,
 the equation (\ref{fritele1}) can be expressed in terms of Hubble parameter
 \begin{equation}
     \frac{f(H)}{4}-\frac{1}{4} H \frac{df(H)}{dH}=4\pi \rho(H),
     \label{friedprva}
 \end{equation}{}
 which is a convenient form in order to reconstruct a specific $f(T)$ function. Again in order to find an oscillatory solution we will demand the form of $a(t)$:
 \begin{equation}
     a(t)=A(1+c \sin (\omega t)),
 \end{equation}{}
 the formula can be inverted to get $a(H)$:
 \begin{equation}
     a(H)=\frac{2 A \omega ^2\pm \sqrt{4 A^2 \left(c^2-1\right) \omega ^2 \left(H^2+\omega ^2\right)+4 A^2 \omega ^4}}{2 \left(H^2+\omega ^2\right)},
     \label{skalaplusminus}
 \end{equation}{}
 then in the matter and radiation dominated era the field equation becomes
 \begin{equation}
      \frac{f(H)}{4}-\frac{1}{4} H \frac{df(H)}{dH}=4\pi \Big(\frac{\rho_m}{a(H)^3}+\frac{\rho_r}{a(H)^4}\Big).
 \end{equation}{}
 For simplicity let us start by taking the parameters $\omega=1$, $\rho_r=0$ and $c=1$ together with (\ref{skalaplusminus}). The  Big Crunch (i.e. oscillations in $a(t)$ with a singularity in $H(t)$ and $R(t)$) periodic scenario solution is then given by:
 \begin{equation}
     f(H)=const H-\frac{2 \pi  H \left(\frac{H^5}{5}+H^3+3 H-\frac{1}{H}\right) \rho_m }{A^3},
 \end{equation}{}
 \begin{equation}
     f(T)=\frac{2 \pi  \rho_m}{{5 A^3}}  \left(\frac{T^3}{1728}-\frac{5 T^2}{144}+\frac{5 T}{4}+5\right) + const \sqrt{-T}.
 \end{equation}{}
 where in the radiation case $\rho_m=0$:
 \begin{equation}
     f(T)=\frac{\rho_r \pi}{35 A^4}   \left(-\frac{5 T^4}{20736}+\frac{7 T^3}{432}-\frac{35 T^2}{72}+\frac{35 T}{3}+35\right) + const \sqrt{-T}.
 \end{equation}{}
 Those are the solutions with $a(t)=A(1+\sin(\omega t))$, where $const \sqrt{-T}$ is again the symmetric part of the lagrangian and in this case is the homogeneous  solution. Interestingly, the cosmological constant is needed to satisfy the oscillatory solution, while the higher order terms are suppressed with respect to the linear term, which must be the case in order to be consistent with the observations -- as the hypothetical corrections to the Teleparalel Einstein-Hilbert action must be  small from the observational point of view. Another important solution is the true cyclic solution, containing no singularities in $H(t)$ and $R(t)$. Again we will use similar parameters as in $f(R)$ analysis, and therefore we set $c=0.8$, $\omega=1$ and $\rho_r=0$ (matter dominated era) assuming $A>0$ and consider the positive sign case in equation (\ref{skalaplusminus}). The solution appears to be rather complicated, but can still be tracked analytically.
 The reconstructed $f(H)$ and $f(T)$ for this case are:
 \begin{multline}
     f(H)=\frac{1000}{729 A^3} \pi  \rho_m  \Big(270 H^2-\sqrt{16-9 H^2} \left(9 H^2+182\right)\\
     -594 H \arcsin {\frac{3 H}{4}}+730\Big)+const H
     \label{matsolh1}
 \end{multline}{}
 \begin{multline}
    f(T)= \frac{1000}{729 A^3} \pi  \rho_m  \Big(-\frac{45 T}{2}+\frac{1}{8} (3 T-728) \sqrt{3 T+64}\\
    -99 \sqrt{3} \sqrt{-T} \arcsin \left(\frac{1}{8} \sqrt{3} \sqrt{-T}\right)+730\Big)+ const \sqrt{-T},
    \label{matsol1}
 \end{multline}{}
 the alternative solution with the negative sign case (\ref{skalaplusminus}) assuming $A>0$ gives the solution:
 \begin{multline}
     f(H)=\frac{1000}{729 A^3} \pi  \rho_m  \Big(270 H^2+\sqrt{16-9 H^2} \left(9 H^2+182\right)\\
     +594 H \arcsin {\frac{3 H}{4}}+730\Big)+const H.
     \label{matsolh2}
 \end{multline}{}
 One of the important properties of this solutions is the symmetric nature of cases assuming $A<0$ with the negative sign in (\ref{skalaplusminus}), and $A>0$ with the positive sign in (\ref{skalaplusminus}), as well as vice versa -- which are equivalent to each other. This property comes from the time inversion symmetry, $t \rightarrow -t$, contained in the corresponding $a(t)$ form, since from $\dot{a}=aH$ follows that $H\rightarrow -H$ if $a\rightarrow a$, under the time inversion. Then, under this inversion, the left side of  (\ref{friedprva}) leads to:
 \begin{equation}
       f(H)-Hf_H=f(-H)-H\frac{df(-H)}{dH},
     \label{uvjet0}
 \end{equation}{}
 which comes from the fact that $T=-12H^2$. Therefore all $f(T)$ theories leading to such solutions are even functions in $H$. This is an important property of cyclic solutions with a symmetric $a(t)$, since there must always exist two connected branches in the dynamical analysis of $a(H)$, as will be discussed later.
 The two $f(T)$ actions for two different scenarios with $c=0.8$ and $c=1$ can seem rather similar, but at the qualitative level the difference is considerable. The crucial point is that the scenario with $c=1$ is not a true cyclic Universe but at some point the Hubble parameter becomes infinite, resulting in a future singularity, while for $c=0.8< 1$
the scale factor undergoes a smooth transition between eternally oscillating phases.
 This analytic results for $f(T)$ gravity will serve as a guidance in deriving and analysing the cyclic behaviour in the context of dynamical system analysis in $f(T)$ cosmology.
 
 \subsection{Dynamical system analysis in $f(T)$ cosmology}
 
 Generally, by exploring the $f(T)$ equations in cosmology with matter fields, it follows that they appear as first order
 differential equations -- since the energy density and pressure contributions contain explicitly the scale factor as a variable. Similar to already discussed dynamic dark energy models, cyclic solutions are still possible in $f(T)$ cosmology, even if it represents only a first order differential system, by virtue of the existence of two connected branches of the phase portrait. Thus the only equation needed is:
 \begin{equation}
       \frac{f(H)}{4}-\frac{1}{4} H \frac{df(H)}{dH}=4\pi \rho(a),
       \label{sacuvanjetorz}
 \end{equation}{}
 where the Hubble parameter is determined by:
 \begin{equation}
     \dot{a}=aH.
     \label{skalaH}
 \end{equation}{}
 The second equation
 \begin{equation}
     \dot{H}=\frac{1}{2}\frac{4\pi \rho(a) (1 + w)}{24 H^2 f_{TT}-f_T}=12\frac{{4\pi \rho(a) (1 + w)}}{f_{HH}}.
     \label{friedtorz2}
 \end{equation}
 can also be obtained from the gravitational field equations, where the fact that $w=p/\rho$ was exploited. However, this equation is simply a time derivative of (\ref{sacuvanjetorz}). By dividing equations (\ref{sacuvanjetorz}) and (\ref{friedtorz2}) one can obtain a new resulting equation 
 \begin{equation}
     \dot{H}=3(1+w)\frac{f-Hf_H}{f_{HH}},
     \label{htocka}
 \end{equation}{}
 which is a one dimensional differential system in terms of the Hubble paramter. However, we will not inspect this equation as it does not contain the information of the dynamics of the scale factor and the interplay between the scale factor and the Hubble parameter -- which is the crucial point in the whole story of the cyclic cosmology. The simple reason for this is that the fixed point corresponding to $\dot{H}=0$ is not of primary importance in the cyclic evolution, but such points are rather given by the minima and maxima of the scale factor, which makes the equation (\ref{sacuvanjetorz}) more appropriate for the analysis in such context. 
 \\ \\
 The equation (\ref{sacuvanjetorz}) is in fact directly determining a phase portrait of $a(H)$. Firstly we are interested in finding the extremal points of $a(H)$, i.e. the points at which the condition $da/dH=0$ must be fulfilled. By differentiating equation (\ref{sacuvanjetorz}) with respect to $a$ we get:
 \begin{equation}
     \frac{da}{dH}=-16\pi Hf_{HH}\Big({\frac{d\rho}{da}}\Big)^{-1}=0,
 \end{equation}{}
 which is fulfilled only if $H=0$ or $f_{HH}(H^*)=0$. As discussed earlier in detail, in cyclic cosmology there must exist two points in time corresponding to minimum and maximum of the scale factor -- the bounce and the turnaround, which are both characterized by $H=0$. Therefore, while discussing cyclic cosmologies the fixed points corresponding to $H=0$ are of central interest. Going to the second derivative we get:
 \begin{equation}
     \frac{d^2a}{dH^2}=-16\pi \Big({\frac{d\rho}{da}}\Big)^{-1}  \Big( f_{HH}  +Hf_{HHH}- Hf_{HH}\frac{da}{dH}\Big),
     \label{druga}
 \end{equation}{}
 where the last term is by definition zero in a fixed point. Since cyclic Universe needs to have both minimum and maximum points of the scale factor at $H=0$, the right-hand side of the equation (\ref{druga}) needs to be both positive - at the time of the cosmological bounce, and negative - at the time of turnaround. This means that there should exist two different branches of $f(H)$ function, and having different signs of second derivative with respect to $H$ at $H=0$. This can also be understood from equations (\ref{sacuvanjetorz}) and (\ref{skalaH}) as those equations will lead to different $f(H)$ branches when $H<0$ and $H>0$.
 To conclude, we thus have:
 \begin{equation}
     \frac{d^2a}{dH^2}=-16\pi \Big({\frac{d\rho}{da}}\Big)^{-1}   f_{HH}(0)
>0, 
     \label{uvjetminmax1}
     \end{equation}
     at the bounce,
     \begin{equation}
         \frac{d^2a}{dH^2}=-16\pi \Big({\frac{d\rho}{da}}\Big)^{-1}   f_{HH}(0)
<0,  
 \end{equation}{}
 at the turnaround. \\ \\
 The similar logic can then also be applied for another fixed point, $H^*$:
 \begin{equation}
      \frac{d^2a}{dH^2}=-16\pi \Big({\frac{d\rho}{da}}\Big)^{-1}   H^*f_{HHH}(H^*)\gtrless 0.
      \label{uvjetminmax2}
 \end{equation}{}
   It is thus obvious that there should exist two different solutions of $f(H)$, where each one must represent a different branch in the phase portrait with the property:
  \begin{equation}
      f_{1HH}(0)=-C_1f_{2HH}(0),
      \label{uvjetminmax11}
  \end{equation}{}
  or
 \begin{equation}
      H^*f_{1HHH}(H^*)=-C_2H^*f_{2HHH}(H^*).
      \label{uvjetminmax22}
  \end{equation}{}
 where $C_1$ and $C_2$ are positive constants given by the specific model function $f(T)$. 
 \\ \\
 Let us assume that for some matter fields the energy density can be represented as $\rho(a)\sim1 /a^n$, then
 the phase portrait can be written in the following form:
 \begin{equation}
     a(H)=c\sqrt[-n]{f-Hf_{H}},
 \end{equation}{}
 the extremal points are:
 \begin{equation}
     \frac{da}{dH}=\frac{Hf_{HH}}{n}(f-Hf_{H})^{-\frac{1}{n}-1}=0,
 \end{equation}{}
 and the second derivative:
 \begin{equation}
     \frac{d^2a}{dH^2}=\frac{1}{n}(f-Hf_{H})^{-\frac{1}{n}-1}(Hf_{HHH}+f_{HH}).
 \end{equation}{}
 In this specific framework of matter fields (which assumes $\rho(a)\sim 1/a^n$ and thus also covers the cases of matter and radiation described as the ideal fluid) it appears that $H=H^*$ with the property $f(H^*)=H^* f_{H}(H^*)$ cannot be a solution as the second derivative vanishes, but it will be demonstrated later that this point has another physical meaning. The only fixed point in this scenario is with $H=0$.
 Then the condition (\ref{uvjetminmax11}) reads:
 \begin{equation}
     f_1(0)^{-\frac{1}{n}-1}f_{1HH}(0)=-C_1f_2(0)^{-\frac{1}{n}-1}f_{2HH}(0).
     \label{uvjetminmax111}
 \end{equation}{}

To conclude, in the case of $f(T)$ gravity we can then establish the following claim: if $f(T)$ is a differentiable function at $T=0$, if $a(t)$ is symmetric in the expanding and contracting phase (i.e if by choosing the bounce as the origin of time, $t_{bounce}=0$, $a(t)=a(-t)$) and if  $H^*=0$ is a fixed point of the system then there always exist a second branch of the $f(T)$ function which gives a symmetric solution in the sense that these two functions are dynamically indistinguishable. This claim can be proven by noting that if there exist a $f(T)$function for which $a(t)$ is an oscillatory solution of the modified Friedman equation in the $f(T)$ theory, than from (\ref{uvjetminmax1}) at the same point the $f(T)$ function must have a maximum and minimum. This can only be satisfied if there is another function $f_2(T)$ satisfying the  property given by (\ref{uvjetminmax11}). From the dynamical perspective those functions are indistinguishable as they give the same $a(t)$ as a solution, even if the phase portraits are different.

\subsubsection{ Concrete examples and applications  in $f(T)$ cosmology}
\begin{figure}[h]
\centering
    \includegraphics[width=0.8\linewidth]{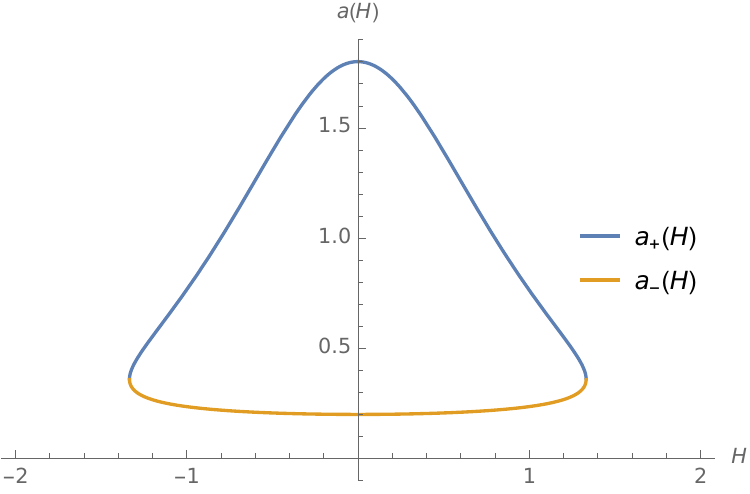}
       \caption{Phase portrait $a(H)$ from (\ref{skalaplusminus}) with parameters $\omega=1$, $A=1$ and $c=0.8$.} \label{fazniah}
\end{figure}

 As a starting point, the analytical solutions (\ref{matsolh1}) and (\ref{matsolh2}), corresponding to solutions $a(t)=A(1+c \sin \omega t)$, can be used to explore the consistency of the dynamical system
 analysis. We will assume that we know nothing about the solution of this system, the only information we are given is the concrete type of the $f(T)$ or simply $f(H)$ theory, in this case:
 \begin{multline}
     f_+(H)=f_1(H)=\frac{1000}{729 A^3} \pi  \rho_m  \Big(270 H^2-\sqrt{16-9 H^2} \left(9 H^2+182\right)\\
     -594 H \arcsin {\frac{3 H}{4}}+730\Big)+const H
     \label{fplus}
 \end{multline}{}
 and the symmetric part
 \begin{multline}
     f_-(H)=f_2(H)=\frac{1000}{729 A^3} \pi  \rho_m  \Big(270 H^2+\sqrt{16-9 H^2} \left(9 H^2+182\right)\\
     +594 H \arcsin {\frac{3 H}{4}}+730\Big)+const H,
     \label{matsolh2}
 \end{multline}{}
 with $A>0$, $\rho_m>0$ and the matter field guided by the evolution with $n=3$. The first derivatives are:
 \begin{multline}
    f_{+H}(H)= \frac{1000 \pi  \rho_m}{27 A^3 \sqrt{16-9
   H^2}}  \Bigg(H \Big(9 H^2+20 \sqrt{16-9 H^2}-16\Big)\\
   -22 \sqrt{16-9 H^2} \arcsin\left(\frac{3 H}{4}\right)\Bigg)+const,
 \end{multline}{}
 \begin{multline}
    f_{-H}(H)= \frac{1000 \pi  \rho_m }{27 A^3 \sqrt{16-9
   H^2}} \Bigg(H \Big(-9 H^2+20 \sqrt{16-9 H^2}+16\Big)\\
   +22 \sqrt{16-9 H^2} \arcsin\left(\frac{3 H}{4}\right)\Bigg)+const,
 \end{multline}{}
 going to the second derivative:
 \begin{equation}
     f_{+HH}(H)=\frac{2000 \pi \rho_m \left(9 H^2+10 \sqrt{16-9 H^2}-41\right)  }{27 A^3 \sqrt{16-9 H^2}}.
     \label{drugadermax}
 \end{equation}{}
 \begin{equation}
     f_{-HH}(H)=\frac{2000 \pi \rho_m \left(-9 H^2+10 \sqrt{16-9 H^2}+41\right)  }{27 A^3 \sqrt{16-9 H^2}},
     \label{drugaderfmin}
 \end{equation}{}
 now, it is straightforwardly visible that the condition (\ref{uvjetminmax111}) is fulfilled:
 \begin{equation}
     f_{+HH}(0)=-\frac{500 \pi  \rho_m }{27 A^3}\sim -f_{-HH}(0)=-\frac{1500 \pi  \rho_m }{A^3}.
 \end{equation}{}
 By improving the understanding of this problem let us plot the phase portrait $a(H)$ in Fig.(\ref{fazniah}).
 It is clear that the maximum and minimum are given at $H=0$, the turnaround points. However, the interesting detail is also given by the interception point where $a_+$ and $a_-$ are equal.
 This is achieved when $dH/da=0=dH/dt \dot{a}^{-1}$ or the point where $\dot{H}=0$. Then from (\ref{htocka}) follows the condition:
 \begin{equation}
     \frac{f_+(H^*)-H^*f_{+H}(H^*)}{f_{+HH}(H^*)}=\frac{f_-(H^*)-H^*f_{-H}(H^*)}{f_{-HH}(H^*)}=0,
     \label{uvjetjednakosti}
 \end{equation}{}
 with $H^*=H(a^*)$, where $a^*$ satisfies $dH(a^*)/da=0$. As it was pointed before, $H^*$ is exactly the point where $f_{HH}(H^*) \rightarrow \infty$ but $f(H^*)-H^*f_{H}(H^*)$ must be constant (energy density at a specific $H$) with the property:
 \begin{equation}
     f_+(H^*)-H^*f_{+H}(H^*)=f_-(H^*)-H^*f_{-H}(H^*)=16 \pi \rho(H^*)=const.,
     \label{uvjetenergije}
 \end{equation}
 where $H^*$ is the point where $a_+$ and $a_-$ meet each other.
 In this concrete example from the second derivative in (\ref{drugadermax}) and (\ref{drugaderfmin}) follows $H^*=\pm 4/3$, and by combining equations (\ref{fplus})-(\ref{drugaderfmin}) the conditions (\ref{uvjetjednakosti}) and (\ref{uvjetenergije}) are simple to check.

 \section{Non-periodic oscillating cosmologies}
 A natural generalization of the  models so far considered in this work is given by the case in which the Universe undergoes phases of contraction and expansion, but is not periodic in some or all of the the parameters chosen to define a configuration space: $a$, $H$ and $R$. Of course, when the departure from periodic evolution stays small the periodic analysis can still be used as a suitable approximation. For instance, in such type of non-periodic scenarios the values of $a_{min}$ and $a_{max}$ can change during different cosmological cycles. The evolution of such cyclic Universe in the configuration space would not be given by closed trajectories, and would - from the point of view of dynamical system analysis - in the simplest case lead to a fixed points given by spirals, rather than centers. This corresponds to oscillating evolution where the system asymptotically either approaches or is being repelled from a given fixed point. For instance, the value of $a_{max}$ can increase during each cycle and be repelled from an unstable spiral fixed point $a_{min}^{*}$. From the mathematical point of view, the analysis is in this case made simpler, since such fixed points have non-vanishing real parts and are thus covered by Hartman-Grobman theorem. In other words, in such type of scenarios it is sufficient to use the linear dynamical analysis given by (\ref{linear}), where $\lambda$ needs to be complex with non-vanishing real parts. There is then no need to use claims similar to Claim 1 - 3 in order to ensure the character of solutions. The analysis of physically viable non-periodic oscillatory cosmological solutions certainly requires a new detailed study in future. Here we limit ourselves to only few preliminary remarks. Such types of oscillatory solutions, where $a_{max}$ grows significantly in each cycle during time, could in principle very effectively solve the problems of growth of density perturbations and dramatic growth of entropy in the contracting phase, as was recently discussed in \cite{cik12} and \cite{ntc}. Namely, if there is a substantial growth of the scale factor prior to contraction, which is only slightly diminished during the period of contraction, then the Universe can be treated as approximately empty during contraction, which naturally solves the usual problems of contracting phase. Likewise, although the total entropy of the Universe would be growing between the cycles, the entropy that can be measured within the physical horizon of an observer can stay approximately the same or even decrease in time -- due to the large growth of the scale factor which leaves only a small fraction of the Universe in the previous cycle within the horizon of an observer during the new cycle. We leave a general analysis of such models as a goal of a future work.    
 \section{Discussion and conclusion}
As there are still no direct empirical evidence, nor complete theories to properly address the issue of the origin of the Universe, since general relativity  blows-up at the initial singularity it predicts, there are two open alternatives answering this fundamental question. The first one simply assumes that the singularity predicted by general relativity is not a pathological result of a theory which is not properly suited for high-curvature and high-energy regimes of the very early Universe, but a real physical beginning of our Universe. The second one, avoiding the physical and philosophical problems of the singular beginning of the Universe, assumes that the initial singularity will be removed in the full theory of quantum gravity yet to be developed. Since the manifold  observations have confirmed that the Universe is not static but expanding, the alternative of the eternal Universe needs to describe the Universe in which the present epoch of expansion is resulting not from a singular beginning, but from a previous phase of contraction. If this picture is extended to have a complete description of the cosmological history, we naturally arrive at the paradigm of perpetually oscillating Universe. 
In the course of time many concrete models of cyclic and bouncing cosmologies were developed assuming specific modifications of general relativity, specific types of additional entities - like various scalar fields, or specific new theoretical frameworks - like string theory. Many features of the proposed models strongly depend on these concrete (and often quite hypothetical) theoretical assumptions on which they rest. It is therefore not straightforward to see which properties describing the dynamics of these cyclic models are general and more robust and which are just consequences of some potentially very specific and hypothetical theoretical assumptions. \\ \\
The objective of this work was to start the investigation of model-independent and general dynamic properties of cyclic cosmological solutions. In other words, we were interested in properties of oscillating solutions which are universal for large classes of gravity theories. After a more general discussion in the first part of the paper, the presented analysis was then implemented on several types of modified gravity frameworks: $f(R)$ gravity, dynamic dark energy and $f(T)$ gravity. Our interest in these types of theories in this paper comes from several reasons: i) they come as a rather direct and conservative extension of the standard general relativity (in the case of $f(R)$ and $f(T)$ gravity - by a direct generalization of the Lagrangian density), ii) because of the previous reason these frameworks are very suitable as  effective and toy-theory approaches to quantum gravity (i.e. for the regimes where we expect that higher order corrections in curvature, or alternatively torsion, will start to play a significant role), iii) despite the fact that the physical extensions of the standard general relativity are minimal in such theories, at the same time they have a rich structure and it is possible to develop various concrete models within them. \\ \\
We started our analysis by reviewing the typical properties of spacetime geometries shared by different models of cyclic cosmologies.  While doing so we proposed the analysis in the configuration space of $a$, $H$ and $R$ (which can be suitably extended with other coordinates if required by the structure of field equations) as a particularly convenient way of describing the evolution of the cyclic Universe. It turns out that this approach was general enough for the application in the cases of $f(R)$, $f(T)$ and dynamic dark energy models. The discussion in the later part of the paper was even simpler for some specific cases of dynamic dark energy and for $f(T)$ cosmologies in general, due to the fact that the field equations could be cast in the form of a one-dimensional system of differential equations, thus making $a$ and $H$ sufficient variables for the complete description of the phase space.
\\ \\ 
In order to obtain general and qualitative conclusions regarding the dynamics of cyclic cosmologies,
a very efficient approach is given by the analysis of the fixed points of the system of equations guiding the cosmological dynamics. The difficulty in applying the linear stability analysis in this type of problem comes from the fact that we are interested in oscillating solutions, corresponding to closed orbits in the phase plane -- while fixed points associated to such solutions, i.e. eliptic fixed points or centers, are not stable with respect to non-linear corrections. We thus discussed how the linear stability analysis can be supported by the two theorems regarding the existence of non-linear centers, in order to be used in the analysis of cyclic cosmologies. We applied these theorems to the cosmological context given by the equations (\ref{skala})-(\ref{kurva}) and demonstrated that the centers found by the linear stability theory will be stable under non-linear corrections if the symmetry condition, $g(A, H, R)=- g(a, -H, R)$, is satisfied in the case of symmetric cyclic cosmologies with respect to the bounce, or if the conserved quantity, $I(a,H,R)$, which we have introduced and discussed how to construct in the same section, has a local extremum at the fixed point. These results are of interest even in the case where the linear stability analysis is not at all applicable, as can often happen in the modified cosmological equations leading to the systems of three or more dimensions including eigenvalues of the stability matrix with vanishing real parts. This is due to the fact that the existence of the extremal point of the conserved integral $I(a, H, R)$ at the fixed point of the system is sufficient to guarantee that the fixed point is stable. \\ \\
After the general discussion in the first part of the paper, we have first focused on discussion of oscillatory solutions in $f(R)$ gravity. We have obtained the solutions to the eigenvalue problem for the linear stability analysis in general $f(R)$ theory, also discussing the form of the conserved integral $I(a,H,R)$ and the condition for its extremum. In order to analyse a typical example of cyclic solutions in this type of gravity theories, we have numerically obtained a $f(R)$ function leading to a simple cyclic and non-singular form of $a(t)$, $H(t)$ and $R(t)$ and then approximated it with a power-law solution. We have then compared the value of the scale factor corresponding to the fixed point of the system of differential equations with the value of the scale factor at which the conserved integral $I(a,H, R)$ has an extremum -- and found that those values are in an excellent agreement. Furthermore, the oscillatory nature of the considered cosmological solution was manifested in the imaginary eigenvalues of the stability matrix for the linear analysis. 
\\ \\
We have next discussed the application to dynamic dark energy models. We have first considered how such problems can be cast in the form of the system of differential equations (\ref{skala})-(\ref{kurva}), assuming that the cosmological term can be written as $\Lambda(a, R, H)$. Then the already discussed application of the theorem on non-linear centers ensures that in such case a center at the point $a=a^{*}$, $H=0$ and $R=0$ predicted by the linear theory will be sufficient to guarantee the existence of a non-linear center, and thus symmetric cyclic cosmological solutions sufficiently close to this point, if $\Lambda(a,H,R)=\Lambda(a, -H, R)$. Next, we discussed in detail the properties of cyclic models realized in a simple toy-theory of dynamic dark energy given by $\Lambda=\Lambda(a)$. It was discussed how it is possible to realize oscillatory solutions in this setting, even though the corresponding system is only the first order differential system. The existence of cyclic cosmological solutions is possible due to the existence of two branches of solutions in the $a'$-$a$ phase space, corresponding to the expansion (positive branch) and contraction (negative branch). These branches need to connect at two distinct fixed points of the modified Friedmann equation written in the form: $a'(t)=f(a)$, which correspond to the minimal and maximal value of the scale factor. Under these requirements it is possible to have both the transition from the contracting to the expanding phase (the bounce), and the transition from the expanding to the contracting phase (the turnaround). We have further discussed that, from the point of view of stability of these fixed points, they need to behave as semi-stable fixed points, so that for the positive branch, the phase  points in the nearby region of the minimum of the scale factor are repelled from that fixed point, while they need to be attracted towards it in the case of the negative branch -- and reverse being true for the second fixed point at the maximum of $a$. Under these conditions, if the state of the contraction is chosen as the initial condition of the evolution of the Universe, the Universe will tend to turn from contraction into expansion in the course of its evolution, while shortly after entering into the expanding phase it will tend to continue the expansion and move away from the region of the phase space corresponding to the contraction -- until it approaches the second fixed point in future, leading to a new entering into the contraction phase. This configuration can thus enable solutions which are perpetually oscillating between the contracting and expanding phases of the cosmological evolution. After this general discussion, we have demonstrated and confirmed our conclusions discussing several concrete realizations of $\Lambda(a)$ cosmologies. After this, we also considered an example of $\Lambda(R)$ cosmology, which can naturally be discussed from the perspective of $f(R)$ cosmology. We revisited the functional form reconstructed in our previous work on cyclic cosmology in $f(R)$ and demonstrated that at the fixed point of the system the condition of the extremum of the conserved integral, $I(a,H,R)$ is satisfied. Further cases, including dynamic dark energy interacting with the matter sector, as well as $\Lambda(H)$ and $\Lambda(a,H, R)$ models, should be addressed in future studies.    
\\ \\
Finally, we have studied the possibility of cyclic solutions in $f(T)$ gravity. Here, as already pointed out by earlier studies, the conditions for realization of bounce and turnaround are mutually exclusive when investigated through the $a$-$H$ functional dependence. However, we have argued that this does not prevent the existence of oscillatory solutions due to the fact that the nature of this dependence is in fact double-valued. Due to this, there will be two branches of the function $H(a)$ which both need to be included for the proper analysis of the phase plane, and thus the dynamics of the Universe. One branch will contain the description of the bounce and the other one of the turnaround. From the point of view of $f(T)$ functional form this means that there will, for a single oscillatory solution of $a(t)$, exist two different branches of $f(T)$. Both of them lead to the same form of cyclic $a(t)$ when the modified Friedmann equation is solved, and they are in this sense dynamically indistinguishable. However, to have a full description of the phase portrait of the cyclic Universe both branches need to be included, due to the already mentioned double-valued nature of $a(H)$ dependence. We have discussed this necessary properties of $f(T)$ cyclic cosmological solutions both in general, where we derived various conditions that need to be satisfied in order to have cyclic solutions, and also using concrete examples. We have first reconstructed the $f(T)$ functional forms leading to cyclic solutions for vacuum and matter radiation era, and then analysed their dynamics in the following sections, thus confirming our general conclusions. 
\\ \\ 
As both our theoretical and observational knowledge related to very strong gravitational fields, necessary for the proper understanding of the origin of the Universe, currently remains incomplete, we believe that a preferred research program in the cosmology of the early Universe consists in investigating the consequences of general types of theories which represent a minimal and logical extension of currently tested general relativity, being suitable for effectively taking into account quantum corrections.
Our aim in this paper was the application of this research program to the problem of cyclic cosmology as an alternative to the idea of a singular beginning of the Universe. We have demonstrated that cyclic cosmological solutions naturally appear in such broad frameworks of modified gravity theories and analysed their general properties. In this sense, our work is complementary with respect to most of earlier studies, which were focused on realizing a specific type of a cyclic cosmology within a concrete theory. Such earlier approach has a clear disadvantage that it depends on the concrete, mostly untested and hypothetically assumed ingredients and modifications of gravity theory, and it is therefore not easy to see which properties of cyclic cosmologies are universal and which particular. \\ \\
One of the necessary future steps in the further development of the described research program of cyclic cosmology would be to focus on the detailed analysis of the special class of viable cyclic models and their properties. This should especially be oriented towards such frameworks where the $\Lambda$CDM phase can be naturally incorporated within the cyclic evolution (see fig.2), and the stability issues of the contracting phase can be solved. Furthermore, it would be necessary to systematically confront generalized classes of cyclic models with observations and predictions of standard cosmology. This could also be done in a more general setting, as the theoretical analysis -- like the one presented in this work -- could be used to obtain the corrections to the standard cosmology supporting cyclic solutions. The parameters describing such corrections should then be systematically confronted with the range of parameters allowed by the available data.   

\section*{Acknowledgement}
Authors would like to thank Kumar Ravi for reading the manuscript and suggestions regarding the graphical representations of cyclic cosmologies.

\end{document}